\documentclass[11pt]{article}
\usepackage{graphicx}
\usepackage{psfrag,cite,amsmath,graphicx,epsfig,latexsym,subfigure,color}
\usepackage[scriptsize,bf]{caption}
\usepackage{color,xcolor,multirow}
\usepackage{amssymb, amsthm,bm}
\usepackage[boxruled]{algorithm}
\usepackage{algpseudocode}
\usepackage{setspace}
\usepackage{array}
\usepackage{booktabs}
\usepackage{hyperref}
\usepackage{url}
\usepackage{ulem}

\newtheorem{claim}{Claim}
\usepackage{xparse}
\NewDocumentCommand{\ShowInline}{v}{%
#1%
}
\newtheorem{remark}{Remark}
\newcommand{\bx}{\bold{x}}
\newcommand{\by}{\bold{y}}

\newcommand{\bh}{\bold{h}}
\newcommand{\bp}{\bold{p}}

\def\bblambda{\boldsymbol \lambda}

\newcommand{\bTheta}{\mathbf{\Theta}}

\usepackage[algo2e,linesnumbered,vlined,ruled]{algorithm2e}

\SetCommentSty{mycommfont}

\begin{document}
\title{Learning to Continuously Optimize Wireless Resource In Episodically Dynamic Environment}

\author{Haoran Sun, Wenqiang Pu, Minghe Zhu,  Xiao Fu, Tsung-Hui Chang,  and Mingyi Hong 
\thanks{H. Sun and M. Hong are with Department of Electrical and Computer Engineering, University of Minnesota, Minneapolis, MN 55455, USA. W. Pu is with the Shenzhen Research Institute of Big Data, Shenzhen, China.  M. Zhu and T-H. Chang are with Shenzhen Research Institute of Big Data, The Chinese University of Hong Kong, Shenzhen, China.   X. Fu is with the School of Electrical Engineering and Computer Science, Oregon State University, Corvallis, OR 97331, USA. Fu is supported by NSF/Intel MLWiNS: CNS-2003082 and M. Hong is supported by NSF/Intel MLWiNS: CNS-2003033.\newline A short version of this paper \cite{sun2021} has been submitted to the 2021 IEEE International Conference on Acoustics, Speech, \& Signal Processing (ICASSP 2021)  on October 21, 2020.}
}
\date{Submitted to ICASSP 2021 on October 21, 2020}

\maketitle
\begin{abstract}
There has been a growing interest in developing data-driven and in particular deep neural network (DNN) based methods for modern communication tasks. For a few popular tasks such as power control, beamforming, and MIMO detection, these methods achieve state-of-the-art performance while requiring less computational efforts, less channel state information (CSI), etc.  However, it is often challenging for these approaches to learn in a dynamic environment where parameters such as CSIs keep changing. 

This work develops a methodology that enables data-driven methods to continuously learn and optimize in a dynamic environment. Specifically, we consider an ``episodically dynamic" setting where the environment changes in ``episodes", and in each episode the environment is stationary.  We propose to build the notion of continual learning (CL) into the modeling process of learning wireless systems, so that the learning model can incrementally adapt to the new episodes, {\it without forgetting} knowledge learned from the previous episodes. Our design is based on a novel min-max formulation which ensures certain ``fairness"  across different data samples. We demonstrate the effectiveness of the CL approach by customizing it to two popular DNN based models (one for power control and one for beamforming), and testing using both synthetic and real data sets.  These numerical results show that the proposed CL approach is not only able to adapt to the new scenarios quickly and seamlessly, but importantly, it maintains high performance over the previously encountered scenarios as well.  
\end{abstract}

\section{Introduction} \label{introduction}
Deep learning has been successful in many applications such as computer vision \cite{voulodimos2018deep}, natural language processing \cite{young2018recent}, and recommender systems \cite{wang2015collaborative}; see \cite{lecun2015deep} for an overview.
 Recent works have also demonstrated that deep learning can be applied in communication systems, either by replacing an individual component in the system (such as signal detection \cite{ye2017power, sun2019deep},  channel decoding \cite{nachmani2018deep}, 
 channel estimation \cite{wen2018deep, sun2018limited}, power allocation \cite{sun2018learning, lee2018deep, liang2019towards,eisen2020optimal,shen2019lorm}, beamforming
 \cite{huang2019fast,zhu2020optimization} and wireless scheduling \cite{cui2019spatial}), or by jointly optimizing the entire system \cite{dorner2017deep, o2017introduction}, for achieving state-of-the-art performance. 
Specifically, deep learning is a data-driven method in which a large amount of training data is used to train a deep neural network (DNN) for a specific task (such as power control). Once trained, such a DNN model will replace conventional algorithms to process data in real time. Existing works have shown that when the real-time data follows similar distribution as the training data, then such an approach can generate high-quality solutions for  non-trivial wireless tasks \cite{ye2017power, sun2019deep,wen2018deep, sun2018limited,sun2018learning,lee2018deep, liang2019towards,eisen2020optimal,zhu2020optimization,shen2019lorm,huang2019fast,nachmani2018deep,cui2019spatial},
while significantly reducing real-time computation, and/or requiring only a subset of channel state information (CSI).  

\noindent{\bf Dynamic environment.} However, it is often challenging to use these DNN based algorithms when the environment (such as CSI and user locations) keeps changing. We list a few reasons below. 

\noindent {\bf 1)} It is well-known that these methods typically suffer from severe performance deterioration when the environment changes, that is, when the real-time data follows a different distribution than those used in the training stage \cite{shen2019lorm}.

\noindent {\bf 2)}  One can adopt the {\it transfer learning} and/or {\it online learning} paradigm, by updating the DNN model according to data generated from the new environment \cite{shen2019lorm}.  However, these approaches usually degrade or even overwrite the previously learned models \cite{parisi2019continual,mccloskey1989catastrophic}. Therefore  they are sensitive to outliers because once adapted to a transient environment/task, its performance on the  existing environment/task can degrade significantly \cite{kirkpatrick2017overcoming}.  Such kind of behavior is particularly undesirable for wireless resource allocation tasks, because the randomness of a typical wireless environment can result in an unstable learning model.

\noindent {\bf 3)} If one chooses to periodically retrain the entire DNN using all the data seen so far \cite{kirkpatrick2017overcoming}, then the training can be {both time and memory} consuming since the number of data needed keeps growing.

Due to these challenges, it is unclear how   state-of-the-art DNN based communication algorithms could properly adapt to new environments quickly without {experiencing} significant performance loss over previously encountered environments. This is one of the main obstacles preventing these data-driven methods {from being} implemented in real communication systems.
 
Ideally, one would like to design  data-driven models that can adapt to the new environment {\it efficiently} (i.e., by using as little resource as possible), {\it seamlessly} (i.e., without knowing when the environment has been changed), {\it quickly} (i.e., adapt well using only a small amount of data), and {\it continuously} (i.e., without forgetting the  previously learned models). Is achieving all the above {goals simultaneously} possible? If so, how to do it? 

\noindent{\bf Continual  Learning.} In the machine learning community, continual learning (CL)  has recently been proposed to address the ``catastrophic forgetting phenomenon", {that is, the tendency of abruptly losing the previously learned models when the current environment information is incorporated \cite{mccloskey1989catastrophic}.} Specifically, consider the setting where different ``tasks" (e.g., different CSI distributions) are revealed sequentially. Then CL aims to retain the knowledge learned from the early tasks through  one of the following mechanisms: 1) regularize the most important parameters 
\cite{kirkpatrick2017overcoming, li2017learning, zenke2017continual}; 2) design dynamic neural network architectures and associate neurons with tasks \cite{rusu2016progressive, yoon2018lifelong}; or 3) introduce a small set of memory for later training rehearsal \cite{lopez2017gradient,shin2017continual,rebuffi2017icarl}. {However, most of the above mentioned methods} require the knowledge of the {\it task boundaries}, that is, the time stamp where an old task terminates and a new task begins. Unfortunately, such a setting does not suit wireless communication problems well, since the wireless environment  usually changes gracefully and swiftly. So it is difficult for wireless systems to identify a precise changing point.  Only limited recent CL works have focused on boundary-free environments \cite{isele2018selective, aljundi2019gradient, rolnick2019experience}, but they both  propose general-purpose tools without considering any problem-specific structures in wireless communications. Further, the effectiveness of these approaches have only been tested on typical machine learning tasks such as image classification. It is unclear whether they will be effective in wireless communication tasks. 

\begin{figure}
  \centering
  \includegraphics[width=0.8\linewidth]{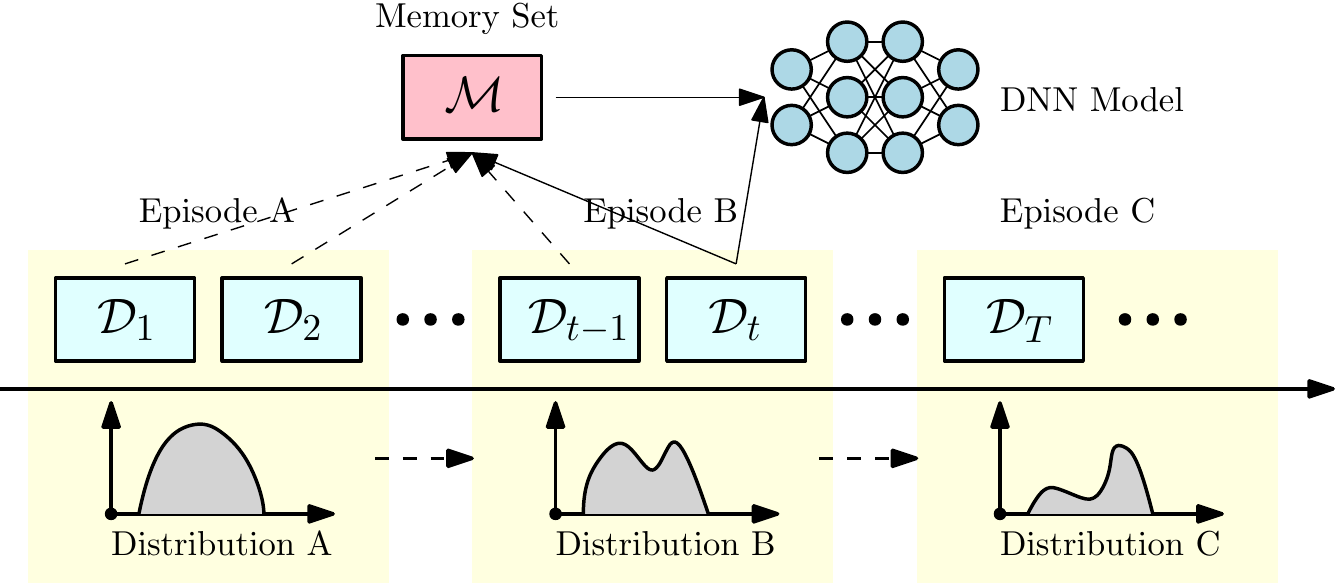}
  \caption{Proposed CL framework for the episodically dynamic environment. The data is feeding in a sequential manner (thus the system can only access $D_t$ at time $t$) with changing episodes and distributions, and the model has a limited memory set $\mathcal{M}$ (which cannot store all data $D_1$ to $D_t$). To maintain the good performance over all experienced data from $D_1$ to $D_t$, the proposed framework optimizes the data-driven model at each time $t$, based on the mixture of the current data $D_t$ and the memory set $\mathcal{M}$. The memory set $\mathcal{M}$ is then updated to incorporate the new data $D_t$. 
  }
  \label{overview}
\end{figure}

\noindent{\bf Contributions.}
The {\it main contribution} of this paper is  that we introduce the notion of CL to data-driven wireless system design, and  develop a  CL formulation together with a training algorithm tailored for core tasks in wireless communications. 
Specifically, we consider an ``episodically dynamic" setting where  the environment changes in {\it episodes}, and within each episodes the distribution of the CSIs stays relatively stationary. 
Our goal is to design a learning model which can seamlessly and efficiently adapt to the changing environment,  without knowing the episode boundary, and without incurring too much performance loss if a previously experienced episode reappears.
We propose a CL framework for wireless systems, which incrementally adapts the DNN models by using data from the new episode as well as a limited but carefully selected subset of data from the previous episodes; see Fig. \ref{overview}.
Compared with the existing {boundary-free} CL  algorithms (which are often general-purpose schemes without a formal optimization formulation) \cite{isele2018selective, aljundi2019gradient, rolnick2019experience}, our
approach comes with a clearly defined and tailored optimization formulation that serves for the wireless resource allocation problem. In particular, our CL method is based on a  min-max formulation which selects a small set of important data samples into the working memory according to a carefully designed {\it data-sample fairness}  criterion. Moreover, we demonstrate the effectiveness of our proposed framework by customizing it to two popular DNN based models (one for power control and the other for multi-user beamforming). We test the proposed CL approach using both synthetic and real data. To advocate the reproducible research, the code of our implementation will be made available online at \url{https://github.com/Haoran-S/ICASSP2021}.

 \section{Literature Review} 
\subsection{Deep learning (DL) for Wireless Communication (WC)}
Recently, deep learning has been used to generate high-quality solutions for non-trivial wireless communication tasks \cite{ye2017power, sun2019deep,wen2018deep, sun2018limited,sun2018learning,lee2018deep, liang2019towards,eisen2020optimal,zhu2020optimization,shen2019lorm,huang2019fast,nachmani2018deep,cui2019spatial}. These approaches can be roughly divided into following two categories: 
\subsubsection{End-to-end Learning}
For the classic resource allocation problems such as power control, the work \cite{sun2018learning} shows that deep neural networks (DNNs) can be exploited to learn the optimization algorithms such as WMMSE \cite{shi2011iteratively}, in an end-to-end fashion. Subsequent works such as \cite{lee2018deep} and \cite{liang2019towards} show that unsupervised learning can be used to further improve the model performance. Different network structures, such as convolutional neural networks (CNN)\cite{lee2018deep} and graph neural networks (GNN) \cite{shen2019graph, eisen2020optimal}, and different modeling techniques, such as reinforcement learning \cite{nasir2019multi}, are also studied in the literature. Nevertheless, all the above mentioned methods belong to the category of end-to-end learning, where a black-box model (typically deep neural network) is applied to  learn either the structure of some existing algorithms, or the optimal solution of a communication task.

\subsubsection{Deep Unfolding}
End-to-end learning based algorithms usually perform well, but their black-box nature makes such kind of models suffer from poor interpretability, i.e., the mechanism of black-box model is usually hard to understand. Further, their performance heavily depends on the quality  and amount of the training samples.  
Alternatively, deep unfolding based methods \cite{balatsoukas2019deep} unfold existing optimization algorithms iteration by iteration and approximate the per-iteration behavior by one layer of the neural  network. The rationale is that, by properly training the network parameters, the multi-layer neural networks can match the behavior of iterative optimization algorithms. The hope is that the trained network uses much fewer iterations (layers), while preserving good interpretability and generalization performance. In the machine learning community, well-known works in this direction include the unfolding of the iterative soft-thresholding algorithm (ISTA) \cite{gregor2010learning}, unfolding of the non-negative matrix factorization methods \cite{hershey2014deep}, and the unfolding of the alternating direction method of multipliers (ADMM) \cite{sprechmann2013supervised}. Recently, the idea of unfolding has been used in communication task  such as MIMO detection \cite{samuel2017deep, samuel2019learning, he2018model}, channel coding \cite{cammerer2017scaling}, resource allocation \cite{eisen2019learning}, channel estimation \cite{borgerding2017amp}, and beamforming problems \cite{hu2020iterative}; see a recent survey paper \cite{balatsoukas2019deep}.

\subsection{Continual Learning (CL)}
\label{section:cl}
{Continual learning is originally proposed to improve reinforcement learning tasks \cite{ring1994continual} to help alleviate the catastrophic forgetting phenomenon, {that is, the tendency of abruptly losing the knowledge about the previously learned task(s) when the current task information is incorporated \cite{mccloskey1989catastrophic}.} It has later been broadly used to improve other machine learning models, and specifically the DNN models; see a recent review article on this topic \cite{parisi2019continual}.} Generally speaking, the CL paradigm can be roughly classified into  the following three main categories.

\subsubsection{Regularization Based Methods} 
Based on the Bayesian theory and inspired by synaptic consolidation in Neuroscience, the regularization based methods penalize the most important parameters to retain the performance on old tasks. Some most popular regularization approaches include Elastic Weight Consolidation (EWC) \cite{kirkpatrick2017overcoming}, Learning without Forgetting (LwF) \cite{li2017learning}, and Synaptic Intelligence (SI) \cite{zenke2017continual}.  However, regularization or penalty based methods naturally introduce trade-offs between the performance of old and new tasks. For example, if a large penalty is applied to prevent the model parameters from moving out of the optimal region of old tasks (i.e., a set of parameters that incur the lowest training/testing cost), the model may be hard to adapt to new tasks. On the other hand, if a small penalty is applied, it may not be sufficient to force the parameters to stay in the optimal region to retain the performance on old tasks. There are also criticisms on the performance loss of regularization based methods for a long chain of tasks \cite{farquhar2018towards,de2019continual}. 

\subsubsection{Architectures Based Methods} By associating neurons with tasks (either explicitly or not), many different types of dynamic neural network architectures are proposed to address the catastrophic forgetting phenomenon \cite{rusu2016progressive, yoon2018lifelong}. However, due to the nature of the parameter isolation, architecture based methods usually require the knowledge of the task boundaries, and thus they are not suitable for wireless settings, where the environment change is often difficult to track. 

\subsubsection{Rehearsal Based Methods} \label{replay-cat} 
Tracing  back  to the 1990s  \cite{robins1995catastrophic},  the  rehearsal  based (aka. memory-based) methods  play an  important  role  in  areas  such  as  reinforcement  learning \cite{mnih2013playing}. As its name suggests, rehearsal based methods use a small set of samples for later training rehearsal, either through selecting and storing the most represented samples such as the incremental classifier, representation learning (iCaRL) \cite{rebuffi2017icarl} and gradient episodic memory (GEM) \cite{lopez2017gradient}, or use generative models to generate representative samples \cite{shin2017continual}. However, both GEM and iCaRL require the knowledge of the task boundaries (i.e., where a new episode starts), which  does  not  suit  wireless  communication problems well. The authors of \cite{rolnick2019experience} propose  boundary-free methods by selecting the samples through random reservoir sampling, which fills the memory set  with data that is uniformly randomly sampled from the streaming  episodes. More complex mechanisms are also introduced recently to further increase the {\it sampling diversity}, where the diversity is measured by either samples' stochastic gradient directions \cite{aljundi2019gradient} or samples' Euclidean distances \cite{hayes2019memory,isele2018selective}.

\subsection{Comparison of CL with other methods}
In this section, we will introduce a few related methods which also deal with streaming data, and  compare them with the CL approach. 

\subsubsection{Online Learning}
As opposed to batch learning, online learning deals with the learning problems where the training data comes sequentially, and data distribution over time may or may not be consistent \cite{shalev2011online}.  
The ultimate goal of online learning is to minimize the cumulative loss over time, utilizing the previously learned knowledge.
In particular, when data sampling is independently and identically distributed, online gradient descent is essentially the stochastic gradient descent method, and all classic complexity results can be applied. On the other hand, when data sampling is non-stationary and drifts over time, online learning methods are more likely to adapt to the most recent data but losing the performance on past data, i.e., suffering from the catastrophic forgetting \cite{aljundi2019online}. 

\subsubsection{Transfer Learning}
Different from online learning, transfer learning is designed to apply the knowledge gained from one problem to another problem, based on the assumption that related tasks will benefit each other, see a complete survey article \cite{pan2009survey}. By transferring the learned knowledge from old tasks to new tasks, transfer learning can quickly adapt to new tasks with fewer samples and less labeling effort. A typical application is the model fine-tuning on a (potentially small) user-specific problem (e.g., MNIST classification) based on some offline pre-trained model using a comprehensive dataset (e.g. ImageNet dataset). By applying the gained knowledge from the original dataset (e.g. ImageNet dataset), the model can adapt to the new dataset  (e.g., MNIST classification) more quickly with fewer samples. Similar ideas have been applied in wireless settings recently \cite{shen2019lorm,yuan2020transfer,zappone2019model} to deal with scenarios that network parameters changes.
However, since the model is purely fine-tuned or optimized  on the new dataset, after the knowledge transfer, the knowledge from the original model may be altered or overwritten, resulting in significant performance deterioration on the original problem \cite{kirkpatrick2017overcoming}.

\begin{figure}
\centering
\begin{minipage}[c]{0.4\linewidth}
    \centering
    \includegraphics[width= \linewidth]{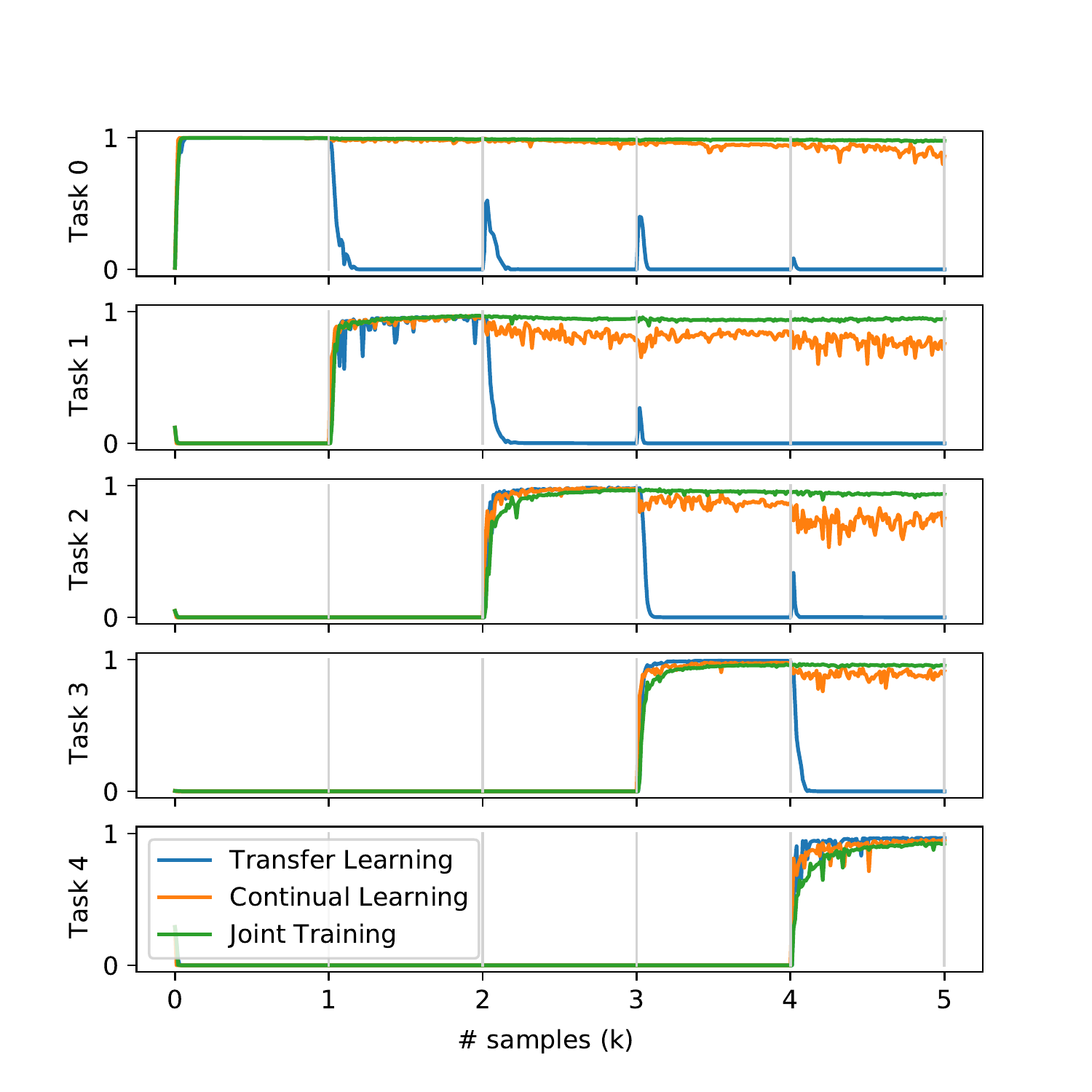}
\end{minipage}
\noindent
\begin{minipage}[c]{0.4\linewidth}
    \centering
    \includegraphics[width= \linewidth]{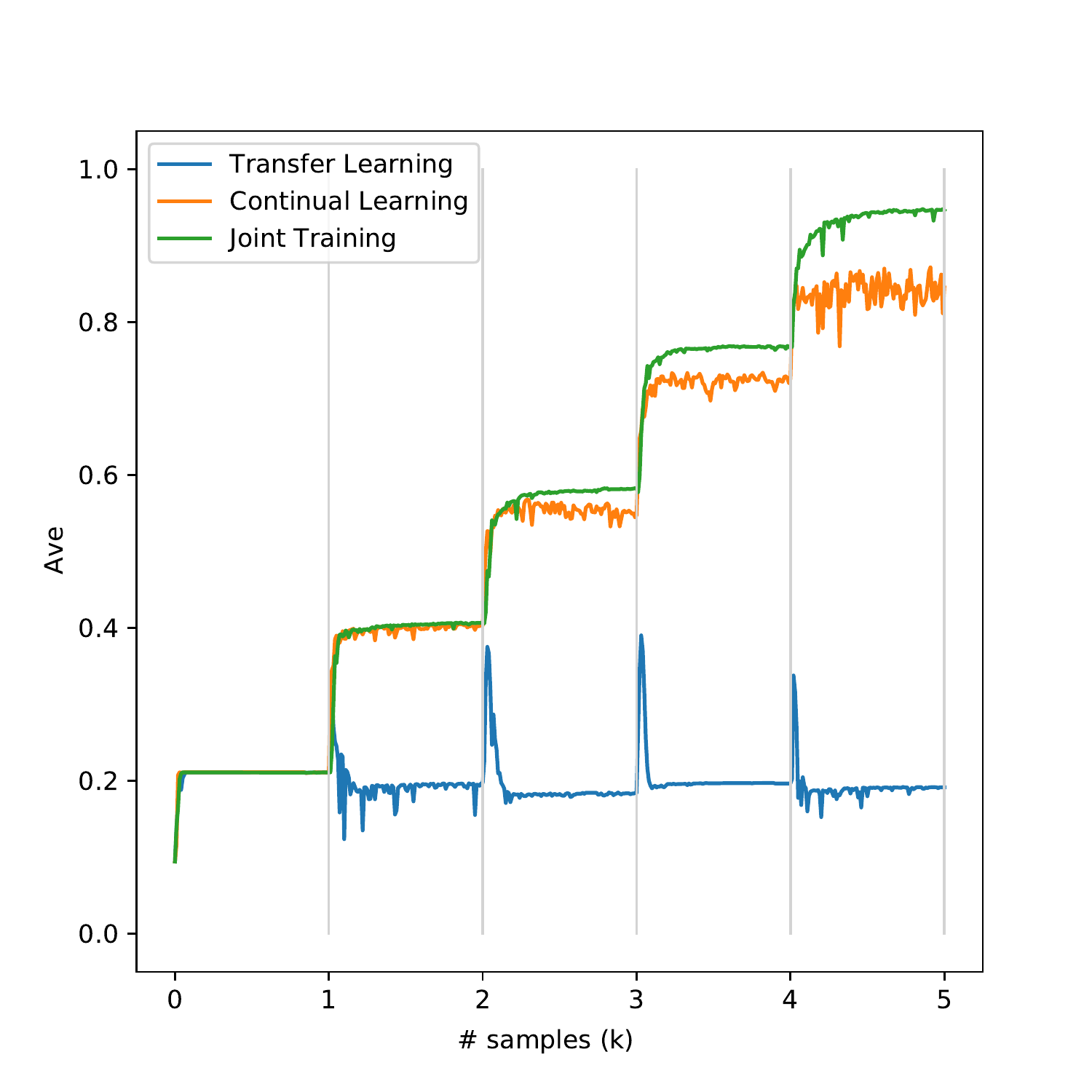}
\end{minipage}
\vspace{-0.4cm}
\caption{MNIST handwritten digits classification example: (left) testing accuracy on each task, which contains two out of ten digits (with no overlapping digits between tasks), (right) testing accuracy on all ten digits. We can observe that naive transfer learning   on one task will significantly degrade the performance on the rest of the tasks, while joint training   using accumulated dataset can preforms well on all experienced tasks.}
\label{mnist_example}
\vspace{-0.4cm}
\end{figure}

\subsubsection{MNIST Example} 
In summary, though being successful in many application scenarios, both transfer learning and online learning are prone to deviate from the original models. We illustrate this phenomenon using the MNIST data set on a handwritten digits classification task. We first split the training data into five disjoint subset, each containing two out of ten digits and $1,000$ samples. We then train a deep neural network model using sequential data (progressively available $10$ samples per time), one task after another. Three training strategies are compared in our simulation, they are transfer learning based on fine-tuning (update the model using previous model as initialization), continual learning (introduce a small memory set for later training rehearsal) and joint training (training from scratch using all accumulated data). Since each time only $10$ samples are available to the model while each task contains $1,000$ samples,  the transfer learning implementation can also be viewed as an online learning method. Testing accuracy on each task and on the entire testing set are reported in Fig. \ref{mnist_example}. One can observe that the naive online learning and/or transfer learning on one task can easily destroy the previously learned parameters and can result in significant performance degradation on the  rest of the tasks, while joint training   using accumulated dataset can preforms well on all experienced tasks. Meanwhile, CL can serve as an efficient alternative to joint training, with much reduced computation and storage overhead.

\section{The Episodic Wireless Environment} \label{formulation}

The focus of this paper is to design learning algorithms in a {\it dynamic} wireless environment, so that the data-driven models we built can seamlessly, efficiently, quickly and continuously adapt to new tasks. This section provides details about our considered {\it dynamic} environment, and discuss potential challenges. 

Specifically, we consider an ``episodically dynamic" setting where the environment changes relatively slowly in ``episodes", and during each episode the learners observe multiple {\it batches} of samples generated from {\it the same} stationary distribution; see Fig. \ref{overview}. We will use $\mathcal{D}_t$ to denote a small batch of data collected at time $t$, and assume that each episode $k$ contains a set of $T_k$ batches, and use  $\mathcal{E}_k = \{\mathcal{D}_t\}_{t\in T_k}$ to denote the data collected in episode $k$. To have a better understanding about the setting of  this work, let us consider the following example. 

\subsection{A Motivating Example}
Suppose a collection of base stations (BSs) run certain DNN based resource allocation algorithm to provide coverage for a given area, e.g., a shopping mall. The users' activities in the area can contain two types of dynamic patterns: 1) regular but gradually changing patterns -- such as regular daily commute for the employees and customers, and such pattern could slowly change from week to week (e.g., the store that people like to visit in the summer is different in winter); 2) irregular but important patterns -- such as large events (e.g., promotion during the anniversary season), during which the distribution of user population (and thus the CSI distribution) will be significantly different compared with their usual distributions, and more careful resource allocation has to be performed.  The episode, in this case, can be defined as ``an unusual period of time that includes a particular event", or a ``usual period of time".

 Suppose that each BS solves a weighted sum-rate (WSR) maximization problem for single-input single-output (SISO) interference channel, with a maximum of $K$ transmitter and receiver pairs.  Let $h_{kk}\in\mathbb{C}$ denote the direct channel between transmitter $k$ and receiver $k$, and $h_{kj}\in\mathbb{C}$ denote the interference channel from  transmitter~$j$ to receiver $k$.  
The power control problem aims to maximize the weighted system throughput via allocating each transmitter's transmit power $p_k$. For a given snapshot of the network, the problem can be formulated as the following:
\begin{align}\label{eq:sum_rate_IA}
\max_{p_1,\ldots, p_K}\quad &R(\bp; \bh):=\sum_{k=1}^K \alpha_k\log\left(1+\frac{|h_{kk}|^2p_k}{\sum_{j\neq k}|h_{kj}|^2p_j+\sigma_k^2}\right) \\
\textrm{s.t.} \quad &0\leq p_k\leq P_{\max},~ \forall \; k=1,2,\ldots, K,\nonumber
\end{align}
where $P_{\max}$ denotes the power budget of each transmitter; $\{\alpha_k>0\}$ are the weights. Problem \eqref{eq:sum_rate_IA} is known to be NP-hard \cite{luo2008dynamic} but can be effectively approximated by many optimization algorithms \cite{shi2011iteratively}. 
The data-driven methods proposed in recent works such as \cite{sun2018learning,lee2018deep, liang2019towards,shen2019lorm,eisen2020optimal} train DNNs using some pre-generated dataset. Here $\mathcal{D}_t$ can include a mini-batch of channels $\{h_{kj}\}$, and each episode can include a period of time where the channel distribution is stationary. 

At the beginning of each week, a DNN model for solving problem \eqref{eq:sum_rate_IA} (pretrained using historical data, say $\mathcal{D}_0$) is preloaded on the BSs to capture the regular patterns in the shopping mall area. Note that one can preload multiple models on each BS, say one model for every six hours. The question is, what should the BSs do  when the unexpected but irregular patterns appear? Specifically, say at 7:00 am a model is loaded to allocate resources up until noon. During this time the BSs can collect batches of data $\mathcal{D}_t, t=1,2, \cdots.$ Then shall the BS update its model during the six hour period to capture the dynamics of the user/demand distribution? If so, shall we use the entire data set, including the historical data and the real-time data, to re-train the neural network (which can be time-consuming), or shall we use transfer learning to adapt to the new environment on the fly (which may result in overwritten the basic 7:00 am model)?

To address the above issues, we propose to adopt the notion of continual learning, so that our model can  incorporate the new data $\mathcal{D}_{t}$ on the fly, while keeping the knowledge acquired from $\mathcal{D}_{0:t-1}$. In the next section, we will detail our proposed CL formulation to achieve such a goal.  

\section{CL for Learning Wireless Resource Allocation} \label{CL_approach}

\subsection{Memory-based CL}

Our proposed method is based upon the notion of the {\it memory-based CL} proposed in \cite{lopez2017gradient,isele2018selective,aljundi2019gradient}, which allows the learner to collect a small subset of historical data for future re-training.  The idea is, once $\mathcal{D}_t$ is received, we  fill in a memory $\mathcal{M}_t$ (with fixed size) with the {\it most representative samples} from all experienced episodes $\mathcal{D}_{0:t-1}$, and then train the neural network at each time $t$ with the data $\mathcal{M}_t\cup\mathcal{D}_t$.  Several major features of this approach are listed below:
\begin{itemize}
    \item The learner does not need to know where a new episode starts (that is, the boundary-free setting) -- it can keep updating $\mathcal{M}_t$ and keep training as data comes in. 
    \item If one can control the size of the memory well (say comparable to the size of each $\mathcal{D}_t$), then the training complexity will be made much smaller than performing a complete training over the entire data set $\mathcal{D}_{0:t}$, and will be comparable with transfer learning approach which uses $\mathcal{D}_t$.
    \item If the size of a given data batch $\mathcal{D}_t$ is very small, the learner is unlikely to {\it overfit} because the memory size is kept as fixed during the entire training process. This makes the algorithm relatively more robust than the transfer learning technique. 
\end{itemize}
 
One of the most intuitive boundary-free memory-based CL methods is the random reservoir sampling algorithm \cite{rolnick2019experience}, which fills $\mathcal{M}_t$ with data uniformly randomly sampled from each of the past episodes.
Other mechanisms include the works that aim to increase the {\it sampling diversity}, and the diversity is either measured by stochastic gradient directions with respect to the sample loss functions \cite{aljundi2019gradient} or samples' Euclidean distances \cite{hayes2019memory,isele2018selective}. However, these works have a number of drawbacks. 
 First, for the reservoir sampling approach, if certain episode only contains a very small number of samples, then samples from this episode will be poorly represented in $\mathcal{M}_t$ because the probability of having samples from an episode in the memory is only dependent on the size of the episode.
Second, for the {diversity based methods}, the approach is again heuristic, since it is not clear how the ``diversity"
measured by large gradient {or Euclidean distances} can be directly linked to the quality of representation of the dataset. Third, and perhaps most importantly, the ways that the memory sets are selected are {\it independent} of the actual learning tasks at hand. This last property makes these algorithms broadly applicable to different kinds of learning tasks, but also prevents them from exploring application-specific structures. It is not clear whether, and how well these approaches will work for the wireless communication applications of interest in this paper.

\newpage
\subsection{The Proposed Approach} \label{section-proposed}
In this work, we propose a new memory-based CL formulation that is customized to the wireless resource allocation problem. 
{Our approach is a departure from the existing memory-based CL approaches discussed in the previous subsection,  because we intend to directly use features of the learning problem at hand to build our memory selection mechanism.}

To explain the proposed approach, let us begin with  presenting two common ways of formulating the training problem for learning optimal wireless resource allocation.
	First, one can adopt an {\it unsupervised learning} approach, which directly optimizes some native measures of wireless system performance, such as the throughput, Quality of Service, or the user fairness \cite{mo2000,hong12survey}, and this approach does not need any labeled data.
	Specifically,  a popular DNN training problem  is given by
\begin{align}\label{eq:unsupervised}
    \min_{\bTheta} 
    \sum_{i\in\mathcal{D}_{0:T}} \ell(\bTheta; \bh^{(i)})
\end{align}
where $\bh^{(i)}$ is the $i$th sample; $\bTheta$ is the DNN weight to be optimized;
$\ell(\cdot)$ is the negative of the per-sample sum-rate function, that is:
$\ell(\bTheta; \bh^{(i)}) = -R(\pi(\bTheta;\bh^{(i)}); \bh^{(i)})$,  where $R$ is defined in \eqref{eq:sum_rate_IA} and 
$\pi(\bTheta;\mathbf{h}^{(i)})$ is the output of DNN which predicts the power allocation. The advantage of this class of unsupervised learning approach is that the system performance measure is directly built into the learning model, while the downside is that this approach can get stuck at low-quality local solutions due to the non-convex nature of DNN \cite{song2021}.  

Secondly, it is also possible to use a supervised learning approach. Towards this end, we can generate some {\it labeled data} by executing a state-of-the-art optimization algorithm over all the training data samples  \cite{sun2018learning}. Specifically, for each CSI vector $\bh^{(i)}$, we can use algorithms such as the WMMSE \cite{shi2011iteratively} to solve problem \eqref{eq:sum_rate_IA} and obtain a high-quality solution $\bp^{(i)}$. Putting the $\bh^{(i)}$ and $\bp^{(i)}$ together yields the $i$th labeled data sample. Such a supervised learning approach typically finds high-quality models \cite{song2021,sun2018learning}, but  often incurs significant computation cost since generating labels can be very time-consuming. Additionally, the quality of the learning model is usually limited by that of label-generating optimization algorithms.

  Our idea is to leverage the advantages of both training approaches to construct a memory-based CL formulation. Specifically, we propose to select the most representative data samples $\bh^{(i)}$'s  into the working memory by using a  {\it sample fairness} criteria, that is, those data samples that have relatively low system performance are more likely to be selected into the memory. Meanwhile, the DNN is trained by performing supervised learning over the selected data samples. Intuitively,  the subset of under-performing data samples is selected to represent a given episode, and we believe that as long as a learning model can perform well on those under-performing samples, then it should perform well for the rest of the samples in a given episode.   

To formulate such a notion of {sample fairness}, 
let us first assume that the entire dataset $\mathcal{D}_{0:T}$ is available. Let us use $\ell(\cdot)$ to denote a function measuring the per-sample training loss, $g(\cdot)$  a loss function measuring system performance for one data sample, $\bTheta$  the weights to be trained, $\bx^{(i)}$  the $i$th data sample and $\by^{(i)}$ the $i$th label. Let $\pi(\bTheta;\bx^{(i)})$ denote the output of the neural network. Let us consider the following bi-level optimization problem  
 \begin{subequations}\label{eq:bilevel}
\begin{align}
\min_{\bTheta}  & \quad \sum_{i \in \mathcal{D}_{0:T}} \lambda^{(i)}_*(\bTheta) \cdot \ell (\bTheta; \bx^{(i)}, \by^{(i)})  \label{eq:bilevel-a}\\
\text{\rm s.t.} & \quad\bblambda_*(\bTheta) = \arg \max_{\bblambda \in \mathcal{B}} \sum_{i \in \mathcal{D}_{0:T}} \lambda^{(i)} \cdot g (\bTheta; \bx^{(i)}, \by^{(i)}), \label{eq:bilevel-b}
\end{align}
\end{subequations}
where $\mathcal{B}$ denotes the simplex constraint
$$\mathcal{B}:= \left\{\bblambda \quad \bigg|  \sum_{i\in \mathcal{D}_{0:T}} \lambda^{(i)} = 1, \quad \lambda^{(i)} \ge 0, \quad  \forall~i\in \mathcal{D}_{0:T}\right\}.$$

In the above formulation, the upper level problem \eqref{eq:bilevel-a} optimizes the {\it weighted} training performance across all data samples, and the lower level problem \eqref{eq:bilevel-b} assigns larger weights to those data samples that have higher loss $g(\cdot)$ (or equivalently, lower system level performance). 
Clearly, the lower level problem has a linear objective value, so the optimal $\bblambda_*$ is always on the boundary of the simplex, and the non-zero elements in $\bblambda_*$ will all have the same weight \footnote{In fact, this problem becomes closely related to the classical min-max fairness criteria; see \cite{lu19TSP-Min-Max} for detailed discussion}. Such a solution naturally selects a subset of data for the upper-level training problem to optimize. A few remarks about the above formulation is in order. 

\begin{remark}
{\bf (Choices of Loss Functions)} One important feature of the above formulation is that we decompose the training problem and the  data selection problem, so that we can have the flexibility of choosing different loss functions according to the applications at hand.  Below we discuss a few suggested choices of the training loss $\ell(\cdot)$ and the system performance loss $g(\cdot)$.

First, the upper layer problem trains the DNN parameters $\bTheta$, so we can adopt any existing training formulation we discussed above. For example, if  supervised learning is used, then one common training loss is the MSE loss:
\begin{align} \label{MSE-loss}
\ell_{\rm MSE}(\bTheta; \bx^{(i)}, \by^{(i)})  & =  \|\by^{(i)} - \pi(\bTheta, \bx^{(i)})\|^2.
\end{align}

Second, we suggest that the lower level loss function $g(\cdot)$  be directly related to the system level performance. For example,
we can set $g(\cdot)$ to be  some adaptive weighted negative sum-rate for $i$th data sample
\begin{align} \label{sumrate-loss}
g_{WSR}(\bTheta, \bx^{(i)}, \by^{(i)}) = -  \alpha_i(\bTheta;\bx^{(i)}, \by^{(i)})\cdot R(\pi(\bTheta, \bx^{(i)}); \bx^{(i)}).
\end{align} 
If we choose $\alpha_i(\bTheta;\bx^{(i)}, \by^{(i)})\equiv 1, \; \forall~i$ , then the channel realization that achieves the worse throughput by the current DNN model will always be selected. However,  this may not be a good choice since the achievable rates at samples across different episodes can vary significantly (e.g., some episodes can have strong interference). Then it is likely that the selected data become concentrated to a few episodes.  
As an alternative, we can choose $ \alpha_i(\bTheta;\bx^{(i)}, \by^{(i)})= \frac{1}{\bar{R}^{(i)}}$, where $\bar{R}^{(i)}$ is the rate achievable by running some existing optimization algorithm on the sample $\bx^{(i)}$. This way, the data samples that achieve the worst sum-rate ``{relative}" to the state-of-the-art optimization algorithm is more likely to be selected. 
Typically the ratio ${1}/{\bar{R}^{(i)}} \cdot R(\pi(\bTheta, \bx^{(i)}); \bx^{(i)})$ is quite uniform  across data samples \cite{sun2018learning}, making it a good indication that a particular data sample is underperforming or not. 
\end{remark}

\begin{remark}
		{\bf (Special Case)} As a special case of problem \eqref{eq:bilevel},  one can choose $\ell(\cdot)$ to be the same as  $g(\cdot)$. 
		Then the bi-level problem reduces to the following min-max problem, which optimizes the {\it worst case} performance (measured by the loss $\ell(\cdot))$) across all samples:
	\begin{align}\label{eq:minmax}
\min_{\bTheta} \max_{\bblambda}  & \quad \sum_{i \in \mathcal{D}_{0:T}} \lambda^{(i)} \cdot \ell (\bTheta; \bx^{(i)}, \by^{(i)})  \\
\text{\rm s.t.} & \quad \sum_{i\in \mathcal{D}_{0:T}} \lambda^{(i)} = 1, \quad 
\lambda^{(i)} \ge 0 \quad  \forall~i\in \mathcal{D}_{0:T}.\nonumber
\end{align}
When $\ell(\cdot)$ is taken as the negative per-sample sum-rate defined in \eqref{eq:unsupervised}, problem \eqref{eq:minmax}  is related to the classical min-max resource allocation  \cite{mo2000,Bengtsson99optimaldownlink,Razaviyayn12maxmin}, with the key difference that it does not achieve {fairness} {\it across users}, but rather to achieve {fairness} across {\it data samples}. 
\end{remark}

\begin{algorithm2e}[tb]
	\DontPrintSemicolon
\caption{Fairness Based Continual Learning Framework}\label{algo_PGD}
   Input: Memory $\mathcal{M}_0=\emptyset$, memory size $M$, max iterations $R$, step-sizes $\alpha$, $\beta$\\
  \While {receive $\mathcal{D}_t$}{
    $\mathcal{G}_t = \mathcal{M}_{t}\cup\mathcal{D}_t$

  \For{$r = 1:R$}{
    $\bTheta  \leftarrow \bTheta - \alpha  \sum_{i \in \mathcal{G}_t}  \lambda_t^{(i)} \cdot \nabla \ell (\bTheta, \bx^{(i)}, \by^{(i)})$\\
    $\bblambda_t   \leftarrow \text{proj} \left(\bblambda_t  + \beta \cdot  G(\bTheta; \bx, \by) \right)$ \tcp*{$G(\bTheta; \bx, \by):= \{g(\bTheta, \bx^{(i)}, \by^{(i)})\}_{i \in \mathcal{G}_t}$} 
}
  \uIf{ $|\mathcal{G}_t|<M$}{
   $\mathcal{M}_{t+1}  = \mathcal{G}_t$.  
   }
  \Else{
   $\text{index} = sort(\bblambda_t)[1:M]$ \tcp*{Pick top $M$ largest weights}
   
   $\mathcal{M}_{t+1}  = \mathcal{G}_t[\text{index}]$ \tcp*{Assign  $M$ samples to  memory}}
}
\end{algorithm2e}

At this point, {neither the bi-level problem \eqref{eq:bilevel} nor the min-max} formulation \eqref{eq:minmax} can be used to design CL strategy yet, because solving these problems requires the full data $\mathcal{D}_{0:T}$. To make these formulations useful for the considered CL setting, we make the following approximation.  Suppose that at $t$-th time instance, we have the memory  $\mathcal{M}_t$ and the new data set $\mathcal{D}_t$ available. Then, we propose to solve the following problem to identify worthy data candidates at time $t$: 
\begin{align}\label{bilevel-ep}
\min_{\bTheta}  & \quad \sum_{i \in \mathcal{M}_t \cup \mathcal{D}_t} \lambda_t^{(i)}(\bTheta) \cdot \ell (\bTheta; \bx^{(i)}, \by^{(i)})   \\
\text{\rm s.t. } &\quad \bblambda_t(\bTheta) = \arg \max_{\bblambda \in \mathcal{B}_t} \sum_{i \in \mathcal{M}_t \cup \mathcal{D}_t} \lambda^{(i)} \cdot g (\bTheta; \bx^{(i)}, \by^{(i)}),  \nonumber
\end{align}
where $\mathcal{B}_t$ denotes the simplex constraint
$$\mathcal{B}_t:= \left\{\bblambda \quad \bigg|  \sum_{i\in \mathcal{M}_t \cup \mathcal{D}_t} \lambda^{(i)} = 1,\quad \lambda^{(i)} \ge 0, \; \forall~i\in \mathcal{M}_t \cup \mathcal{D}_t  \right\}.$$

Similarly, if $\ell(\cdot)$ and $g(\cdot)$ are the same, then the following min-max problem will be solved:
\begin{equation}
\label{minmax-ep}
\begin{aligned}
&\min_{\bTheta} \max_{\bblambda} & & \sum_{i \in \mathcal{M}_t \cup \mathcal{D}_t} \lambda^{(i)} \cdot \ell (\bTheta, \bx^{(i)}, \by^{(i)})  \\
&\text{subject to} & & \sum_{i\in \mathcal{M}_t \cup \mathcal{D}_t } \lambda^{(i)} = 1, \lambda^{(i)} \ge 0, \; \forall~i\in \mathcal{M}_t \cup \mathcal{D}_t.
\end{aligned}
\end{equation}
More specifically, at a given time $t$, the above problem will be {\it approximately} solved, and we will collect $M$ data points $j\in \mathcal{M}_t \cup \mathcal{D}_t$ whose corresponding $\lambda^{(j)}$ are the largest. These data points will form the next memory $\mathcal{M}_{t+1}$, and problem \eqref{bilevel-ep} (or \eqref{minmax-ep}) will be solved again; See Algorithm 1 for details.  Note that in the table, we have defined $\bblambda$ and $G(\bTheta, \bx, \by)$ as concatenations of all local weights and loss functions in $\mathcal{G}_t:= \mathcal{M}_t\cup \mathcal{D}_t$, i.e., $\bblambda:= \{\lambda^{(i)}\}_{i\in \mathcal{M}_t\cup \mathcal{D}_t}$ and
$G(\bTheta; \bx, \by):= \{g(\bTheta, \bx^{(i)}, \by^{(i)})\}_{i\in \mathcal{M}_t\cup \mathcal{D}_t}$, respectively.

 Note that  line 4-6 in Algorithm 1 is motivated by a class of alternating projected gradient descent ascent (PGDA) algorithm recently proposed for solving non-convex/concave min-max problem \eqref{minmax-ep}, see   \cite{razaviyayn2020nonconvex} for a recent survey about related algorithms. 
Such an iteration converges to some stationary solutions for the min-max formulation \eqref{minmax-ep} (see the claim below), while it is not clear if it works for the bi-level formulation \eqref{bilevel-ep}. Nevertheless, we found that in practice Algorithm 1 works well for both formulations.  
\begin{claim} {\cite[Theorem 4.2]{nouiehed2019solving}} Consider the min-max problem \eqref{minmax-ep}  with loss function $\ell(\cdot)$ that continuously differentiable in both $\bTheta$ and $\bblambda$, and  there exists constants  $L_1$ and $L_2$ such that for every $\bTheta_1$, $\bTheta_2$ and data points $\bx, \by$, we have 
\begin{align*}
    \left\| \ell (\bTheta_1, \bx, \by) - \ell (\bTheta_2, \bx, \by) \right\|\le L_1 \left\| \bTheta_1-\bTheta_2 \right\|,  \\
    \left\|   \nabla_{\bTheta} \ell (\bTheta_1, \bx, \by) -   \nabla_{\bTheta} \ell (\bTheta_2, \bx, \by) \right\|\le L_2 \left\| \bTheta_1-\bTheta_2 \right\|.
\end{align*}
For each time instance $t$,  if we  apply the algorithm with the following iterative updates
\begin{align*}
    \bTheta & \leftarrow \bTheta - \alpha  \sum_{i\in \mathcal{M}_t \cup \mathcal{D}_t} \lambda_t^{(i)} \cdot \nabla \ell (\bTheta; \bx^{(i)}, \by^{(i)}),\\
    \bblambda_t  & \leftarrow \text{\rm proj} \left(\bblambda_t  + \beta\cdot    G(\bTheta; \bx, \by) \right),
\end{align*}
then there exists some appropriately chosen step-sizes $\alpha, \beta$ such that the algorithm converges to a stationary point of the problem \eqref{minmax-ep}. 
\end{claim}

\section{Experimental Results} \label{simulation}
In this section, we illustrate the performance of the proposed CL framework using two different applications: 1) power control for weighted sum-rate (WSR) maximization problem with single-input single-output (SISO) interference channel defined in  \eqref{eq:sum_rate_IA}, where the end-to-end learning based DNN is used; see \cite{sun2018learning} for details about the architecture; 2) beamforming for WSR maximization problem  with multi-input single-output (MISO) interference channel,  where deep unfolding based DNN architecture is used; see \cite{zhu2020optimization} for details about the architecture.
By demonstrating the effectiveness of the proposed methods on these two approaches, we believe that our approach can be extended to many other related problems as well.

\subsection{Simulation Setup}
The experiments are conducted on Ubuntu 18.04 with Python 3.6, PyTorch 1.6.0, and MATLAB R2019b on one computer node with two 8-core Intel Haswell processors and 128 GB of memory. The codes are made available online through \url{https://github.com/Haoran-S/ICASSP2021}.

\subsection{Randomly Generated Channel}\label{subsection: Randomly Generated Channel}
We first demonstrate the performance of our proposed framework using randomly generated channels, for a scenario with $K=10$ transmitter-receiver pairs. We choose three standard types of random channels used in previous resource allocation literature \cite{sun2018learning,liang2019towards} stated as following:

\noindent{\bf Rayleigh fading:} Each channel coefficient $h_{ij}$ is generated according to a standard normal distribution, i.e.,  
\begin{align}\label{Rayleigh}
   {\sf Re}(h_{ij})\sim \frac{\mathcal{N}(0, 1)}{\sqrt{2}}, \quad {\sf Im}(h_{ij})\sim \frac{\mathcal{N}(0, 1)}{\sqrt{2}},  \quad \forall i, j \in \mathcal{K}. 
\end{align}
\\\noindent{\bf Rician fading:} Each channel coefficient $h_{ij}$ is generated according to a Gaussian distribution with 0dB $K$-factor, i.e., 
  $${\sf Re}(h_{ij})\sim \frac{1+\mathcal{N}(0, 1)}{2}, \quad {\sf Im}(h_{ij})\sim \frac{1+\mathcal{N}(0, 1)}{2},  \quad \forall i, j \in \mathcal{K}.$$ 
\\\noindent{\bf Geometry channel:}
All transmitters and receivers are uniformly randomly distributed in a R$\times$R area. The channel gains $| h_{ij}|^2$ , which follow the pathloss function in \cite{inaltekin2009unbounded}, can be written as
\begin{align*}
    | h_{ij}|^2= \frac{1}{1+d_{ij}^2} | f_{ij}|^2, \; \forall~i,j
\end{align*}
where $f_{ij}$ is the small-scale fading coefficient follows $\mathcal{CN}(0, 1)$, $d_{ij}$ is the distance between the $i$th transmitter and $j$th receiver.

\begin{figure}
\centering
\begin{minipage}[c]{0.4\linewidth}
    \centering
    \includegraphics[width= \linewidth]{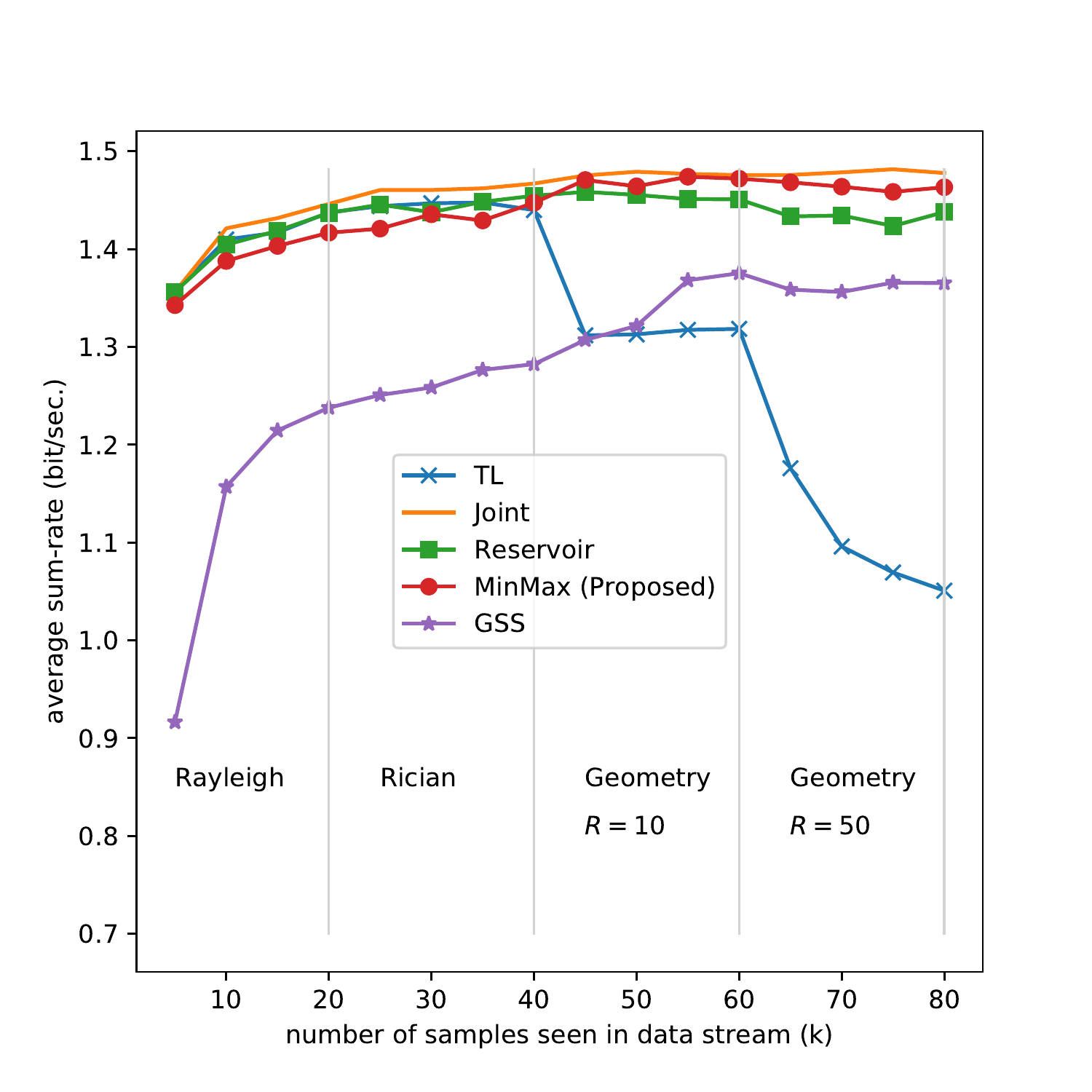}
\end{minipage}
\noindent
\begin{minipage}[c]{0.4\linewidth}
    \centering
    \includegraphics[width= \linewidth]{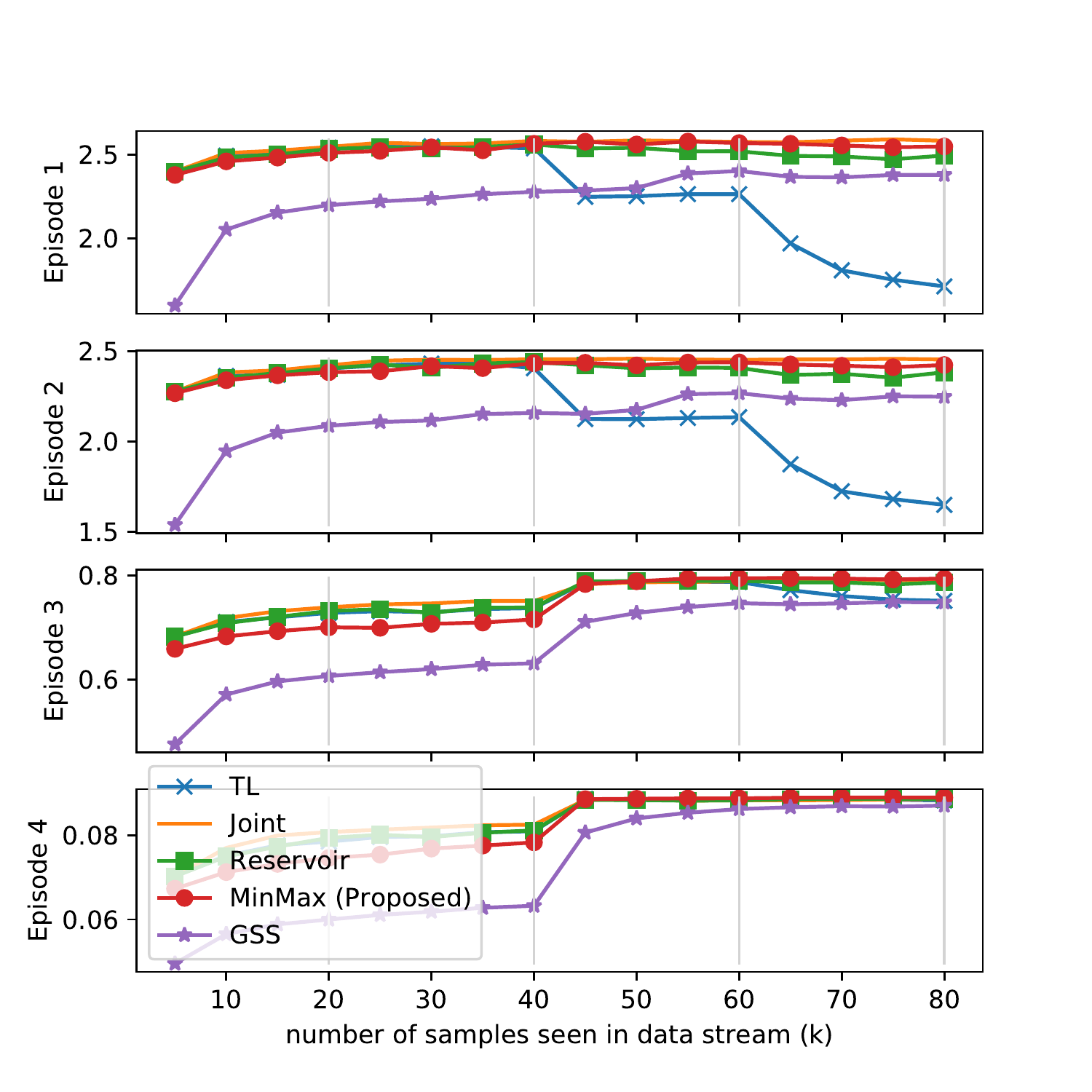}
\end{minipage}
\caption{Testing performance on randomly generated channels. (left) Average sum-rate performance comparison on the mixture test set of all four episodes.  (right) Average sum-rate performance comparison for each individual episode. The gray line indicates the change of episode, which is unknown during training time. }
\label{fig-random-rate}
\end{figure}

Then, we use these coefficients to generate four different episodes as marked in Fig. \ref{fig-random-rate}, namely the Rayleigh fading channel, the Ricean fading channel, and the geometry channel (with nodes distributed in a $10m \times 10m$ and a $50m \times 50 m$ area). Note that this way of generating data and organizing the episodes may not make sense in practice. It is only used at this point as a ``toy" data set. Later we will utilize real data to generate more practical scenarios. For each episode, we generate $20,000$ channel realizations for training and $1,000$ for testing. We also stacked the test data from all episodes to form a mixture test set, i.e., containing $4,000$ channel realizations. During the training stage, a total of $80,000$ channel realizations are available. A batch of $5,000$ realizations is  revealed each time, and the memory space contains only $2,000$ samples from the past. That is, $|\mathcal{D}_{1:16}| = 80,000$, $|\mathcal{D}_{t}|=5,000, |\mathcal{M}_t|=2,000, \; \forall~t$. 

For the data-driven model, we use the end-to-end learning based fully connected neural network model as implemented in  \cite{sun2018learning}. For each data batch of $5,000$ realizations at time $t$, we train the model $\bTheta_t$ using following five different approaches for $100$ epochs, using  previous  model $\bTheta_{t-1}$ as initialization
\begin{enumerate}
    \item Transfer learning (TL) \cite{shen2019lorm} -- update  the  model  using the current data batch (a total of $5,000$ samples);
    
    \item Joint training (Joint) -- update the model using all accumulated data up to current time (up to $80,000$ samples);
    
    \item Reservoir sampling based CL (Reservoir)  \cite{isele2018selective} -- update the model using both the current data batch and the memory set (a total of $7,000$ samples), where data samples in the memory set are uniformly randomly sampled from the streaming episodes;
    
    \item Gradient sample selection  based CL (GSS) \cite{aljundi2019gradient} -- update the model using both the current data batch and the memory set (a total of $7,000$ samples), where data samples in the memory set are selected by maximizing the samples’ stochastic gradient directions diversity;
    
    \item Proposed fairness based CL (MinMax) in Algorithm \ref{algo_PGD}  -- update the model using both the current data batch and the memory set (a total of $7,000$ samples), where data samples in the memory set are selected   according to the proposed data-sample fairness criterion \eqref{bilevel-ep}. Unless otherwise specified, as suggested in Section \ref{section-proposed},  the training loss $\ell(\cdot)$ is chosen as the MSE loss \eqref{MSE-loss}, the  system performance loss $g(\cdot)$ is chosen as the adaptive  weighted  negative  sum-rate loss \eqref{sumrate-loss}, and the weights is chosen as the sum-rate  achievable  by  the WMMSE method \cite{shi2011iteratively}. 
\end{enumerate}

\begin{figure}
\centering
\begin{minipage}[c]{0.4\linewidth}
    \centering
    \includegraphics[width= \linewidth]{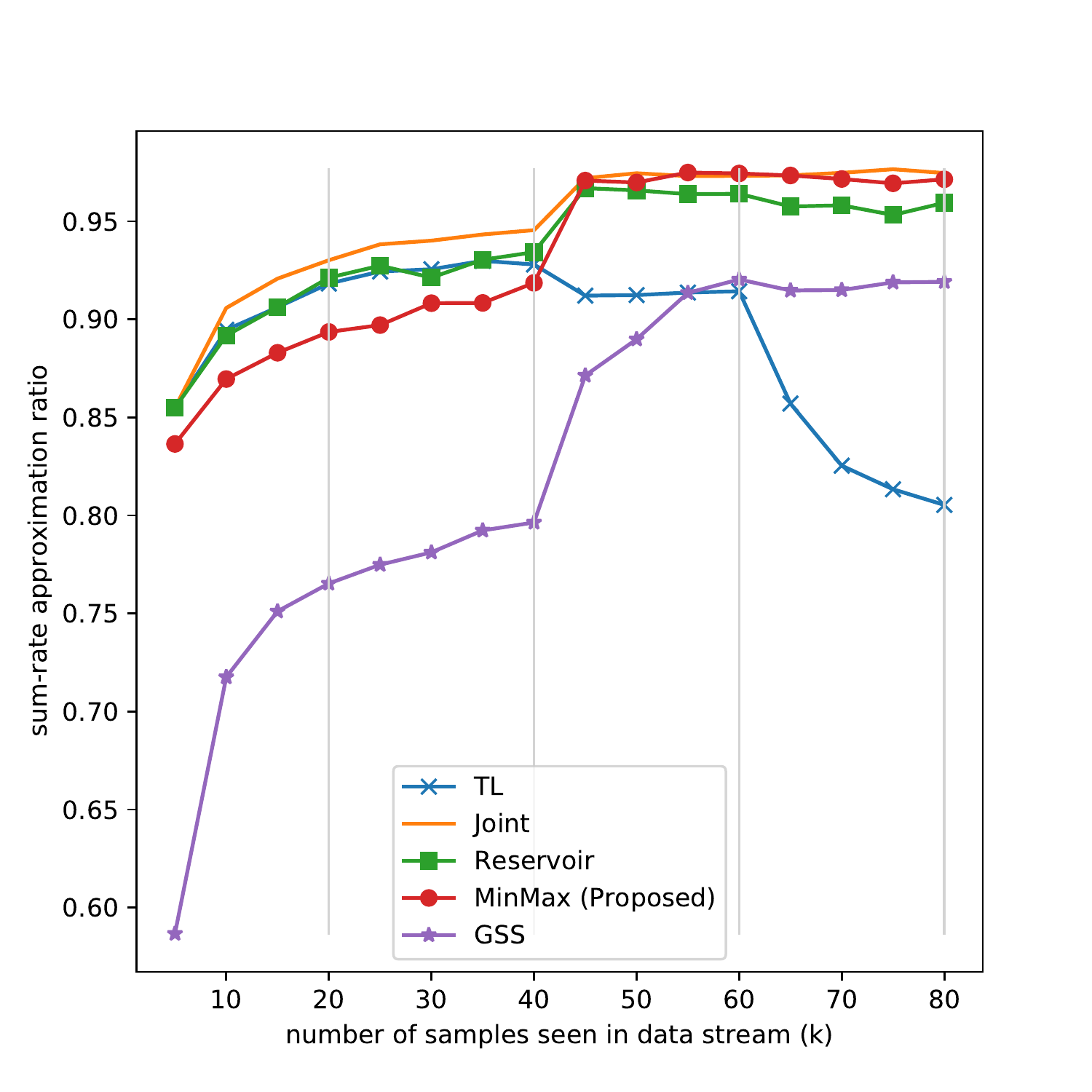}
\end{minipage}
\noindent
\begin{minipage}[c]{0.4\linewidth}
    \centering
    \includegraphics[width= \linewidth]{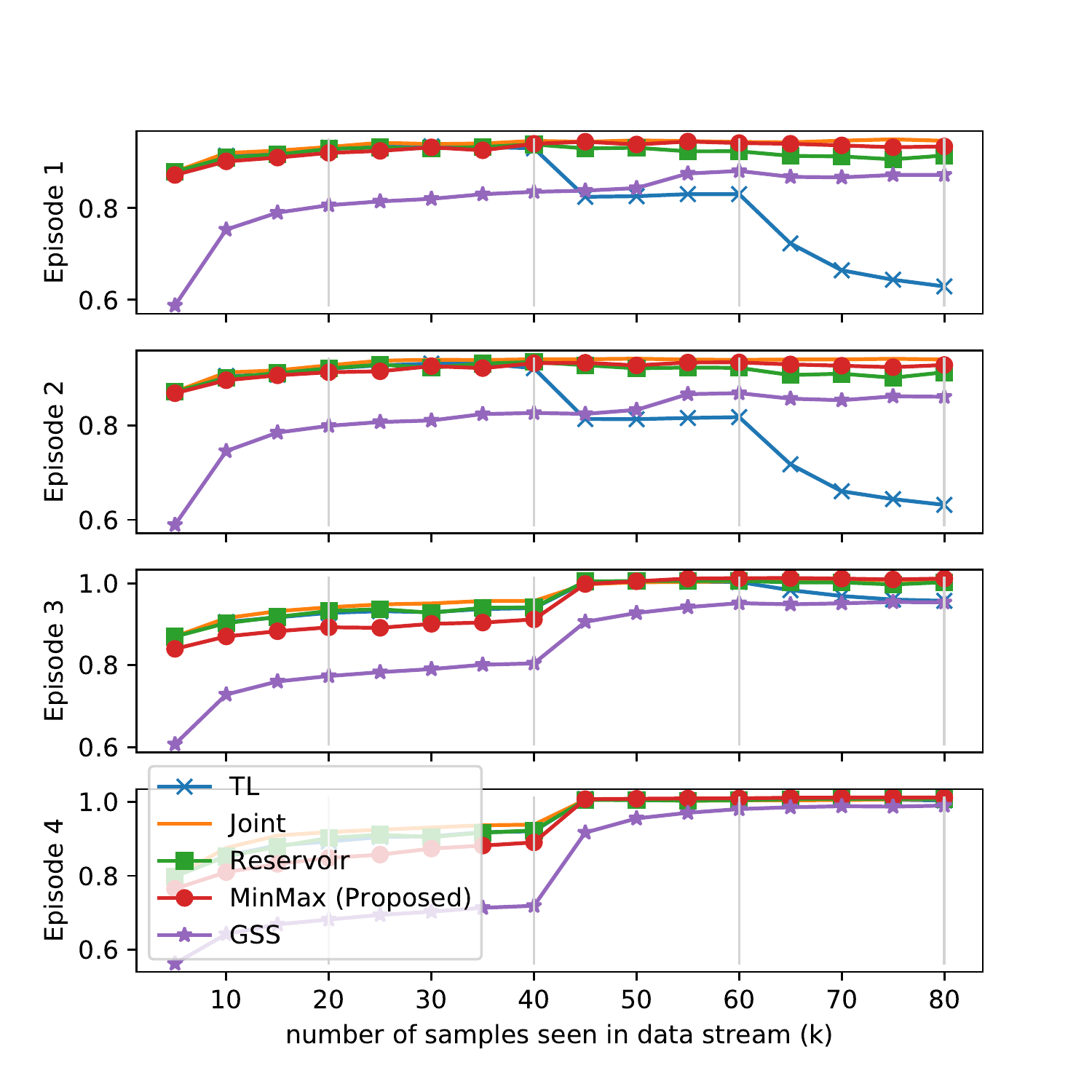}
\end{minipage}
\caption{Testing performance on randomly generated channels. (left) Average sum-rate approximation ratios comparison on the mixture test set of all four episodes.  (right) Average sum-rate approximation ratios comparison for each individual episode. The ratio is computed by the achievable sum-rate of the indicated method divided by the sum-rate achieved by the WMMSE algorithm. The gray line indicates the change of episode, which is unknown during training time.}
\label{fig-random-ratio}
\end{figure}

\begin{figure}
\centering
\begin{minipage}[c]{0.4\linewidth}
    \centering
    \includegraphics[width= \linewidth]{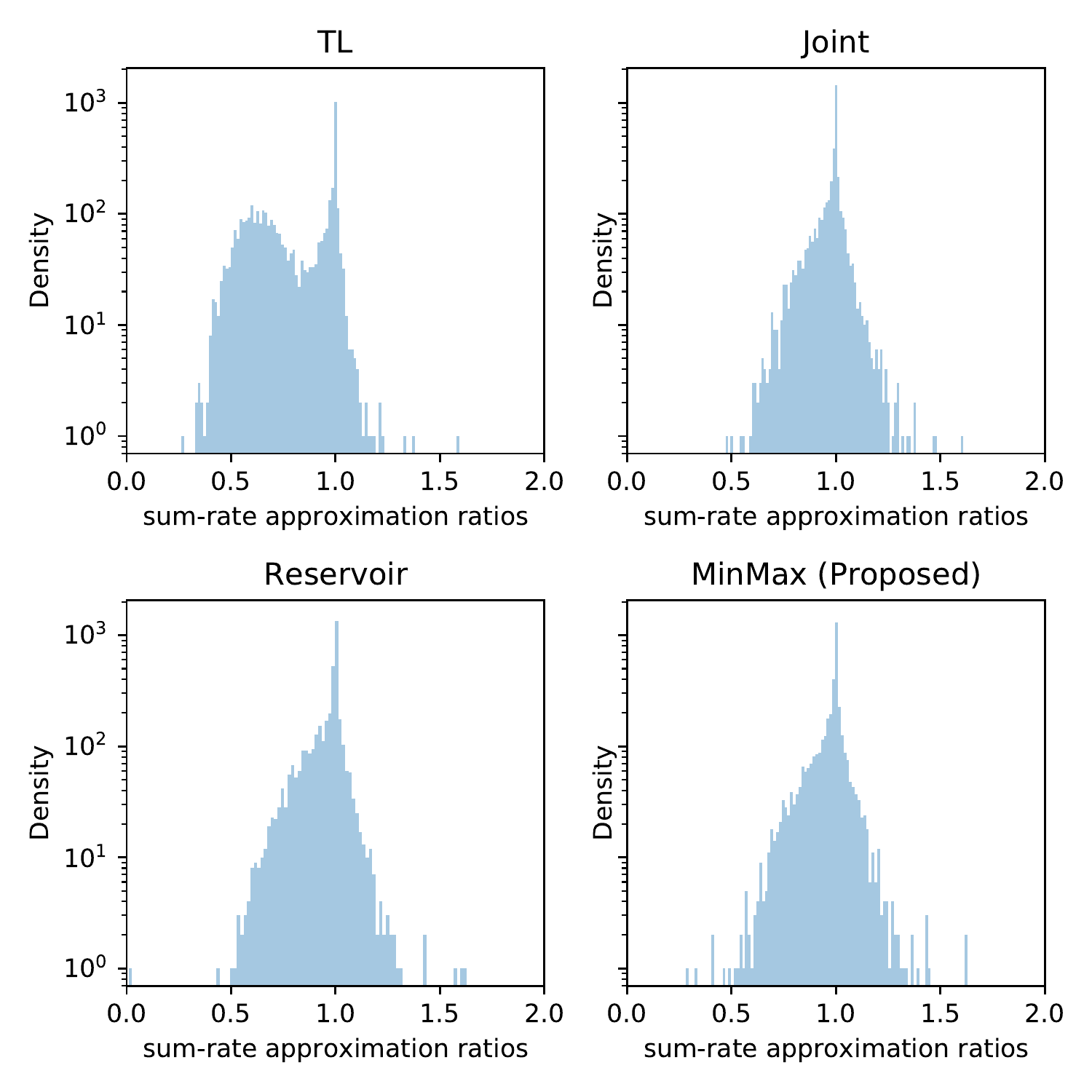}
    \footnotesize{(a) Probability Density Functions (PDF)}
\end{minipage}
\noindent
\begin{minipage}[c]{0.4\linewidth}
    \centering
    \includegraphics[width= \linewidth]{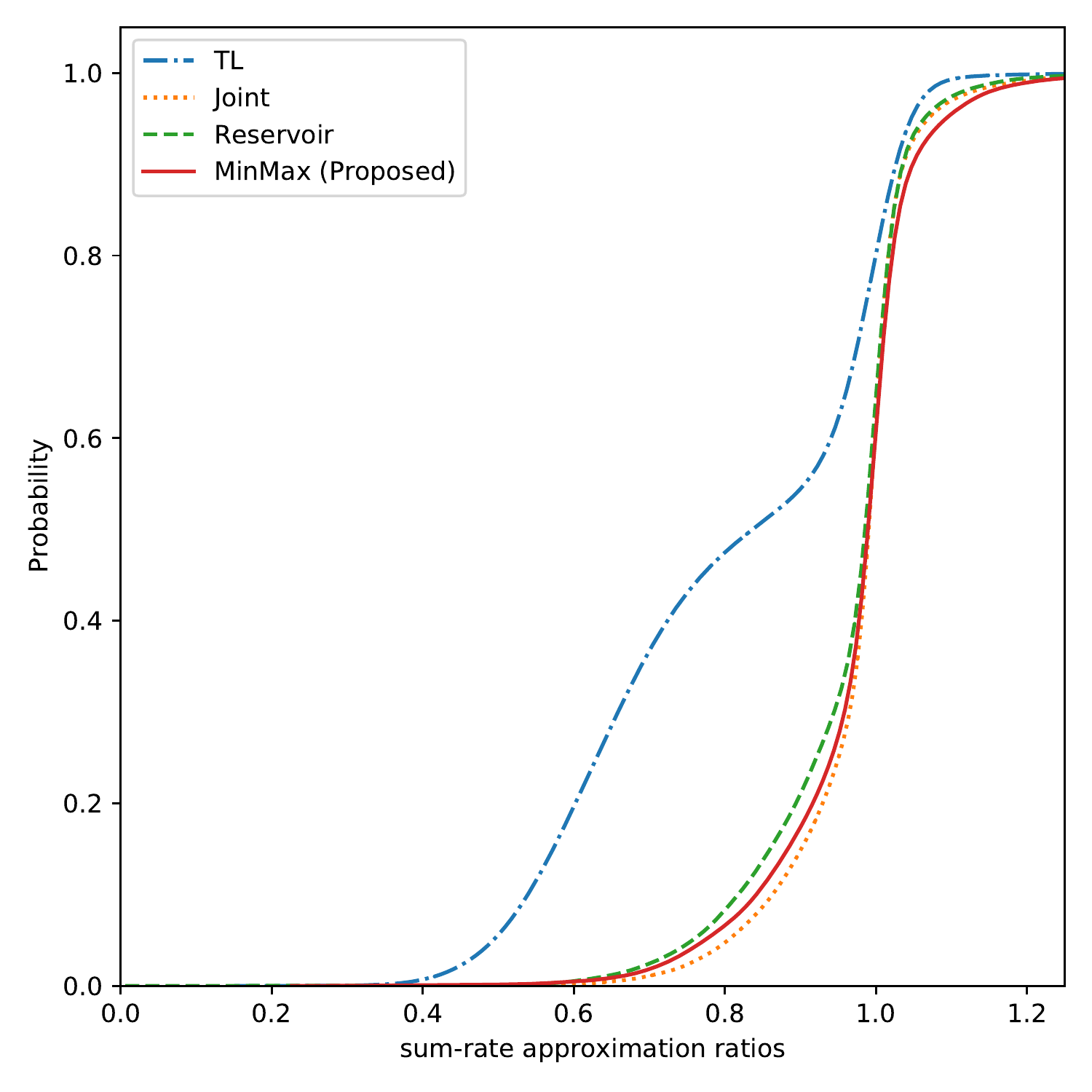}
    \footnotesize{(b) Cumulative Distribution Function (CDF)}
\end{minipage}
\caption{Distribution of the per-sample sum-rate approximation ratio  evaluated at the last time stamp on the mixture test set of all four episodes. The ratio is computed by the achievable sum-rate of the indicated method divided by the sum-rate achieved by the WMMSE algorithm.}
\label{rate_dist}
\end{figure}

The simulation results of five different approaches are compared and shown in Fig.  \ref{fig-random-rate}  - \ref{rate_dist}. In specific,  Fig. \ref{fig-random-rate} shows the trends of the average sum-rate performance, Fig. \ref{fig-random-ratio} shows how the above sum-rate compared with those rates obtained by the WMMSE algorithm \cite{shi2011iteratively}, i.e., approximation ratios, and Fig. \ref{rate_dist} shows the distribution of the approximation ratios for the mixture test set at the last timestamp. Note that the joint training method uses up to $80,000$ in memory spaces and thus violates our memory limitation (i.e., $7,000$ in total), and the transfer learning method adapts the model to new data each time and did not use any additional memory spaces. We can observe that the proposed fairness based CL method performs well over all tasks (cf. Fig. \ref{fig-random-rate} -  \ref{rate_dist}), nearly matching the performance of the joint training method, whereas the transfer learning suffers from some significant performance loss as the ``outlier" episode comes in (i.e., geometry channel in our case). Furthermore, we can observe that the proposed fairness based CL method outperforms other CL-based methods (Reservoir and GSS), in terms of both the average sum-rate (cf. Fig. \ref{fig-random-rate}) and sample fairness (cf. Fig. \ref{rate_dist})  (i.e., fewer samples are distributed in the low sum-rate region) over all tasks. This validates that the proposed approach indeed incorporates the problem structure and advocates fairness across the data samples.      

\begin{figure}
\centering
\begin{minipage}[c]{0.4\linewidth}
    \centering
    \includegraphics[width= \linewidth]{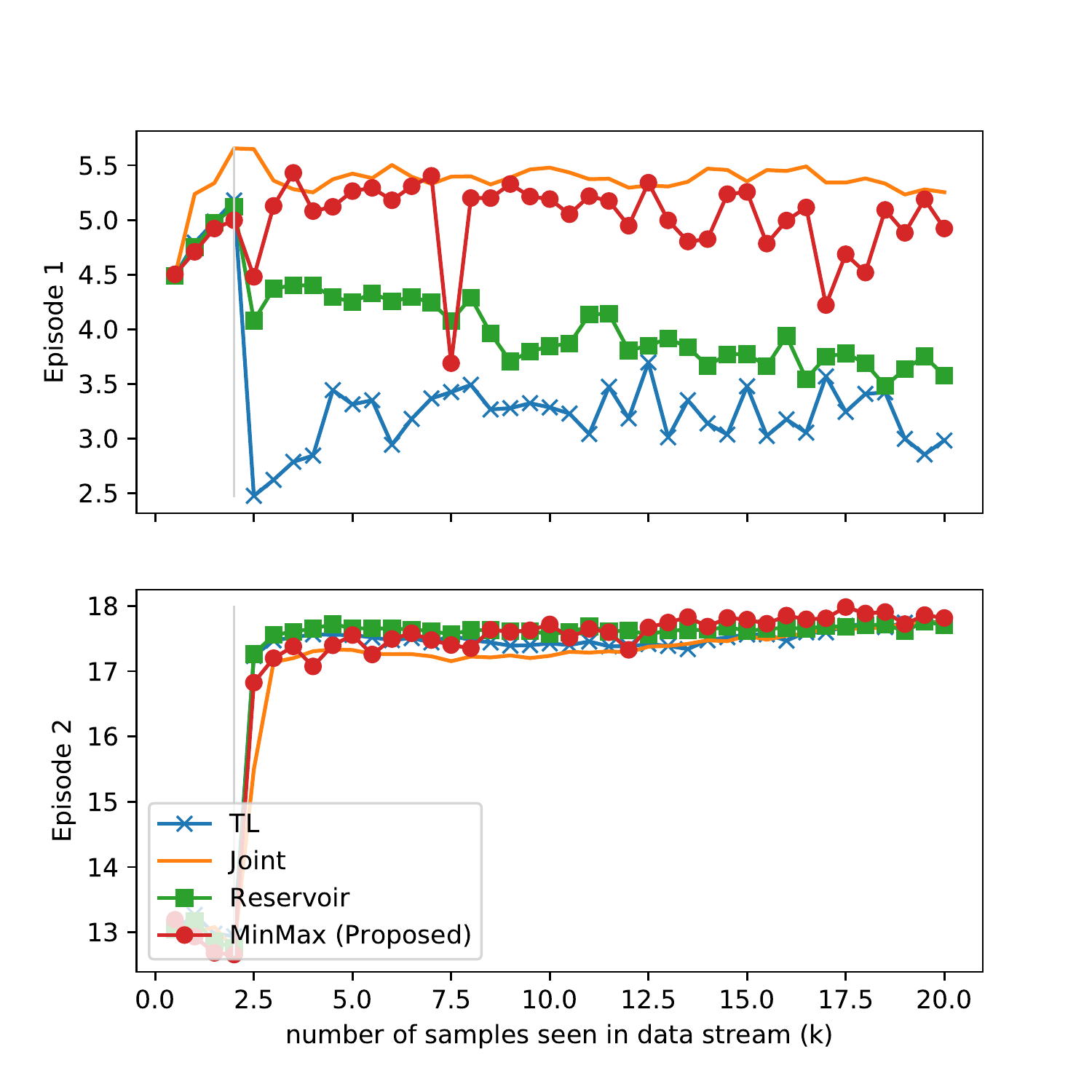}
\end{minipage}
\noindent
\noindent
\begin{minipage}[c]{0.4\linewidth}
    \centering
    \includegraphics[width= \linewidth]{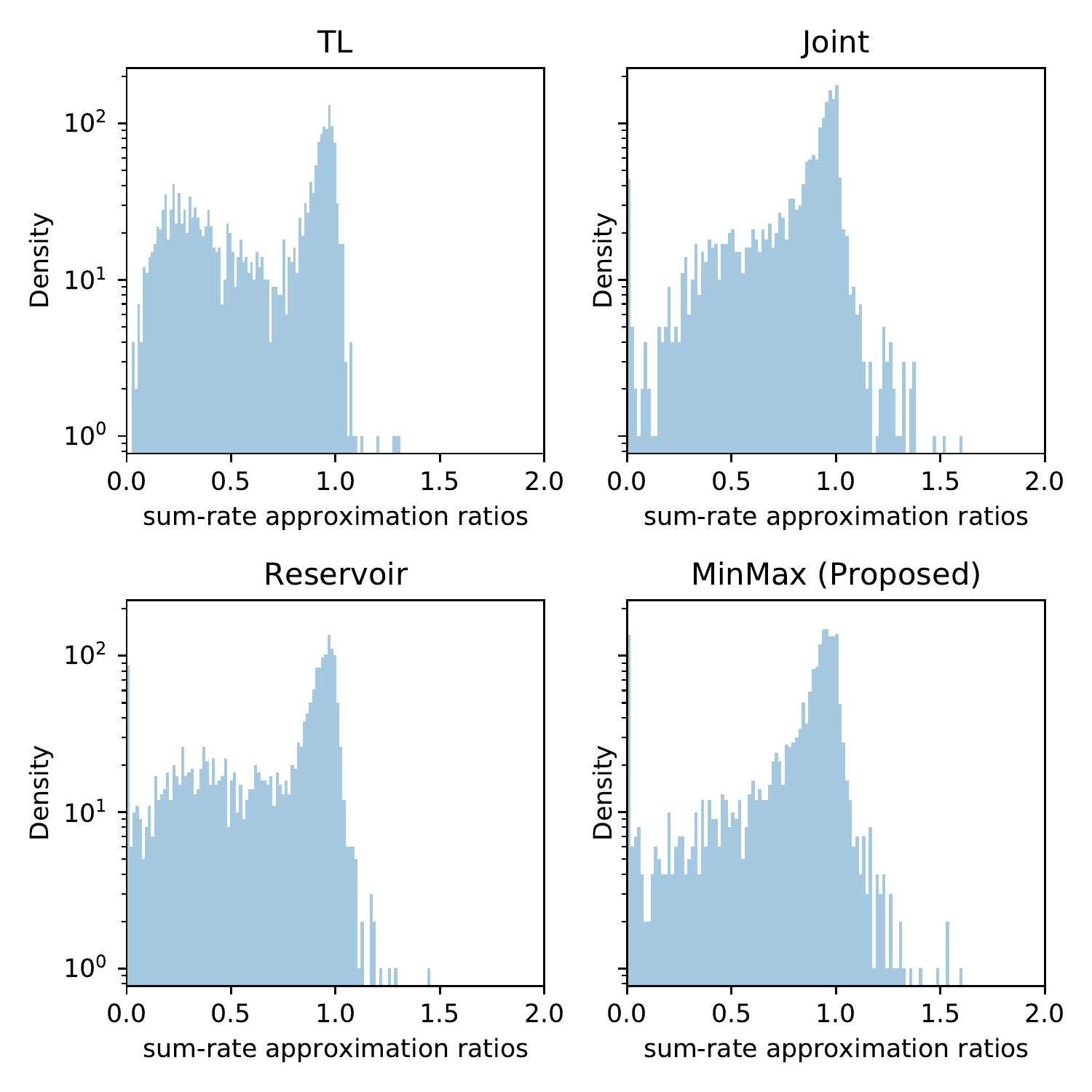}
\end{minipage}
\vspace{-0.4cm}
\caption{Example on unbalanced data (2,000 vs 18,000) and small memory size (0.1k) to showcase of drawbacks of the naive reservoir sampling.  Simulation results on (left) sum-rate and (right) sum-rate ratio compared with WMMSE, show that the reservoir sampling performs similar to transfer learning while the proposed fairness based methods almost matches the best performance achieved by joint training with accumulated data. }
\label{unbalance_example}
\end{figure}

We also demonstrate the drawbacks of the reservoir sampling method using a two-episode unbalanced dataset (each with $2,000$ and $18,000$ samples, respectively) with a small memory size ($100$ samples). We consider two episodes where episode $1$ has $2,000$ samples, and it follows the original Rayleigh fading described in \eqref{Rayleigh};  episode $2$ has $18,000$ samples, and it follows the Rayleigh fading but with strong direct channels (the diagonal entries has magnitude $5$ times larger). Simulation results in Fig. \ref{unbalance_example} on both the sum-rate and the sum-rate ratio (compared with WMMSE) show that the reservoir sampling performs worse in unbalancing scenarios, since the episodes with fewer samples will be poorly represented in the memory set (i.e., memory set contains around $\frac{2,000}{2,000+18,000} \times 100 = 10$ samples from episode 1, out of a $100$ memory size). On the other hand, the proposed fairness based method almost matches the best performance achieved by joint training with accumulated data. 

\subsection{Real Measured Channel}
\begin{figure}
\centering
\begin{minipage}[c]{1\linewidth}
    \centering
    \includegraphics[width=0.5\textwidth]{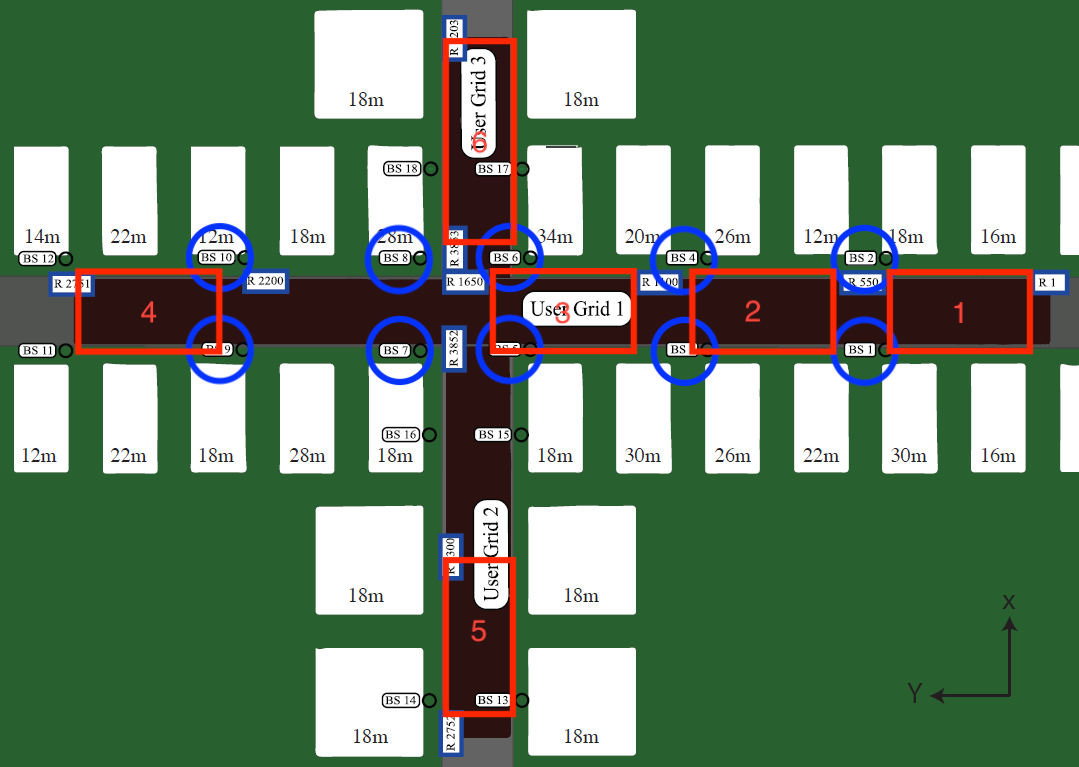}
    \caption{Top view of the DeepMIMO ray-tracing scenario \cite{Alkhateeb2019}, showing the two streets, the buildings, the 10 base stations (blue circle), and the user x-y grids (red rectangular). }
    \label{location}
\end{minipage}
\end{figure}

\begin{figure}
\centering
\begin{minipage}[c]{0.4\linewidth}
    \centering
    \includegraphics[width=\textwidth]{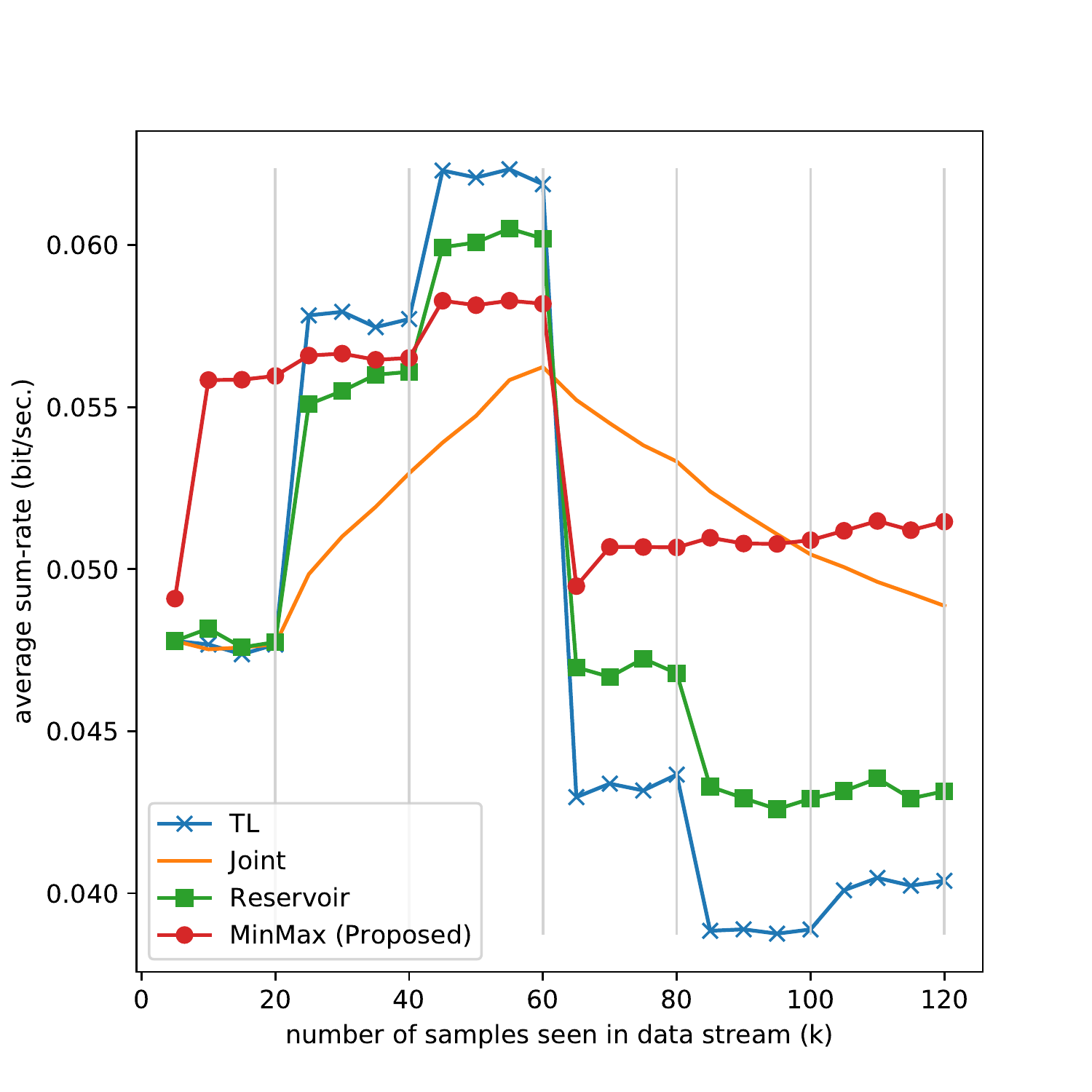}
\end{minipage}
\noindent
\begin{minipage}[c]{0.4\linewidth}
    \centering
    \includegraphics[width=\textwidth]{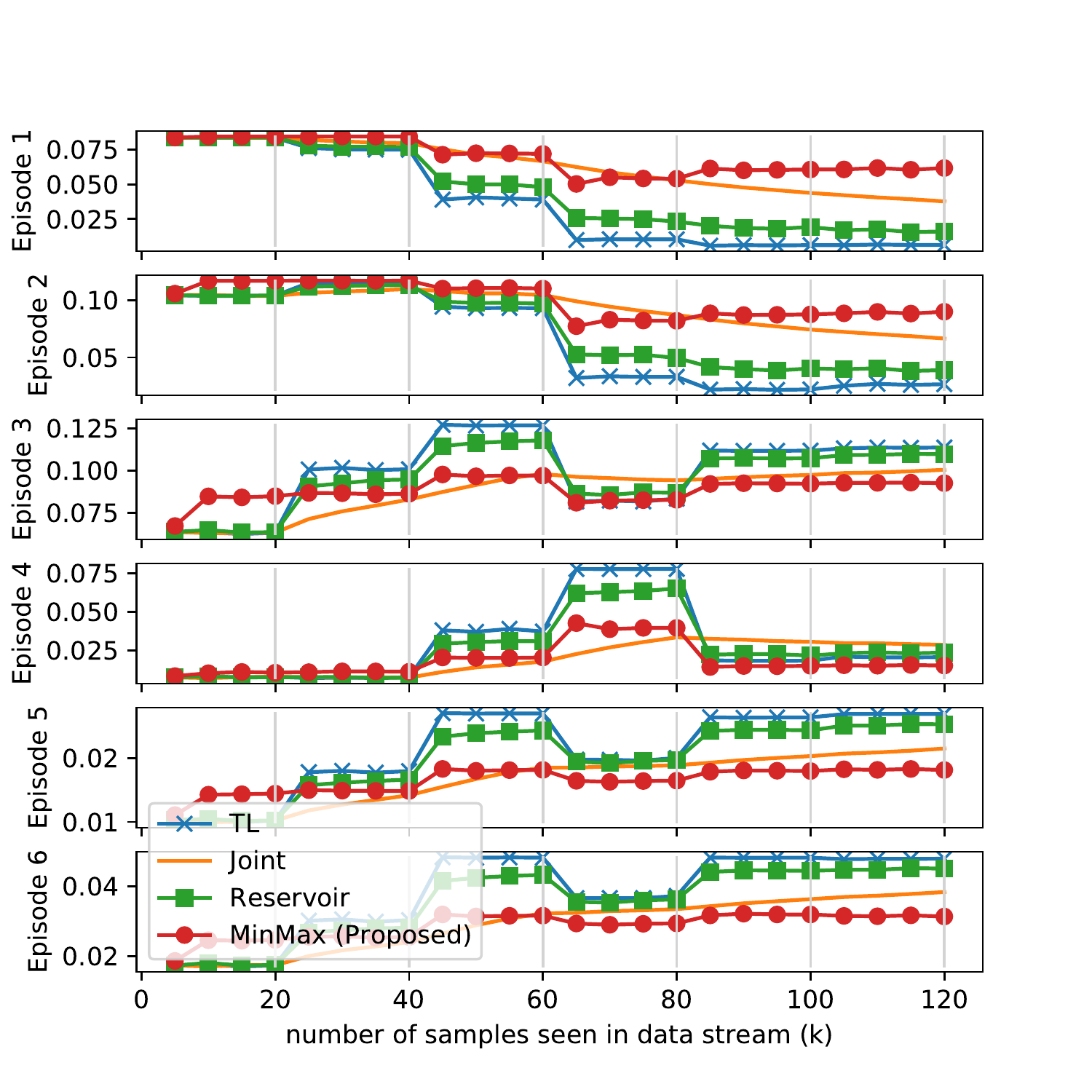}
\end{minipage}
\caption{Average sum-rate performance comparison on real measured channels. (left) performance on mixture of all six episodes. (right) per-task sum-rate performance. }
\label{rate_pertask_mimo}
\end{figure}

\begin{figure}
\centering
\begin{minipage}[c]{0.4\linewidth}
    \centering
    \includegraphics[width=\textwidth]{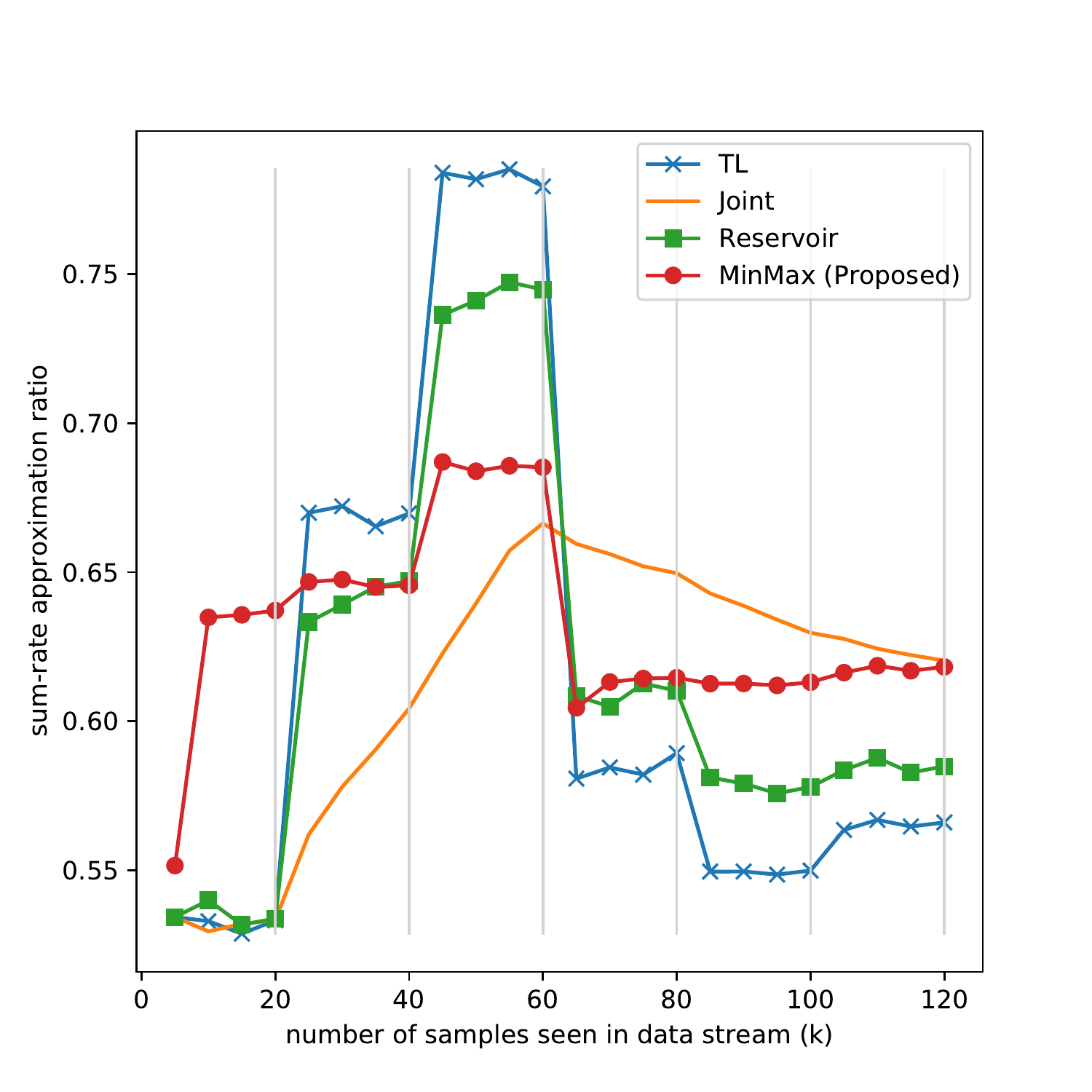}
\end{minipage}
\noindent
\begin{minipage}[c]{0.4\linewidth}
    \centering
    \includegraphics[width=\textwidth]{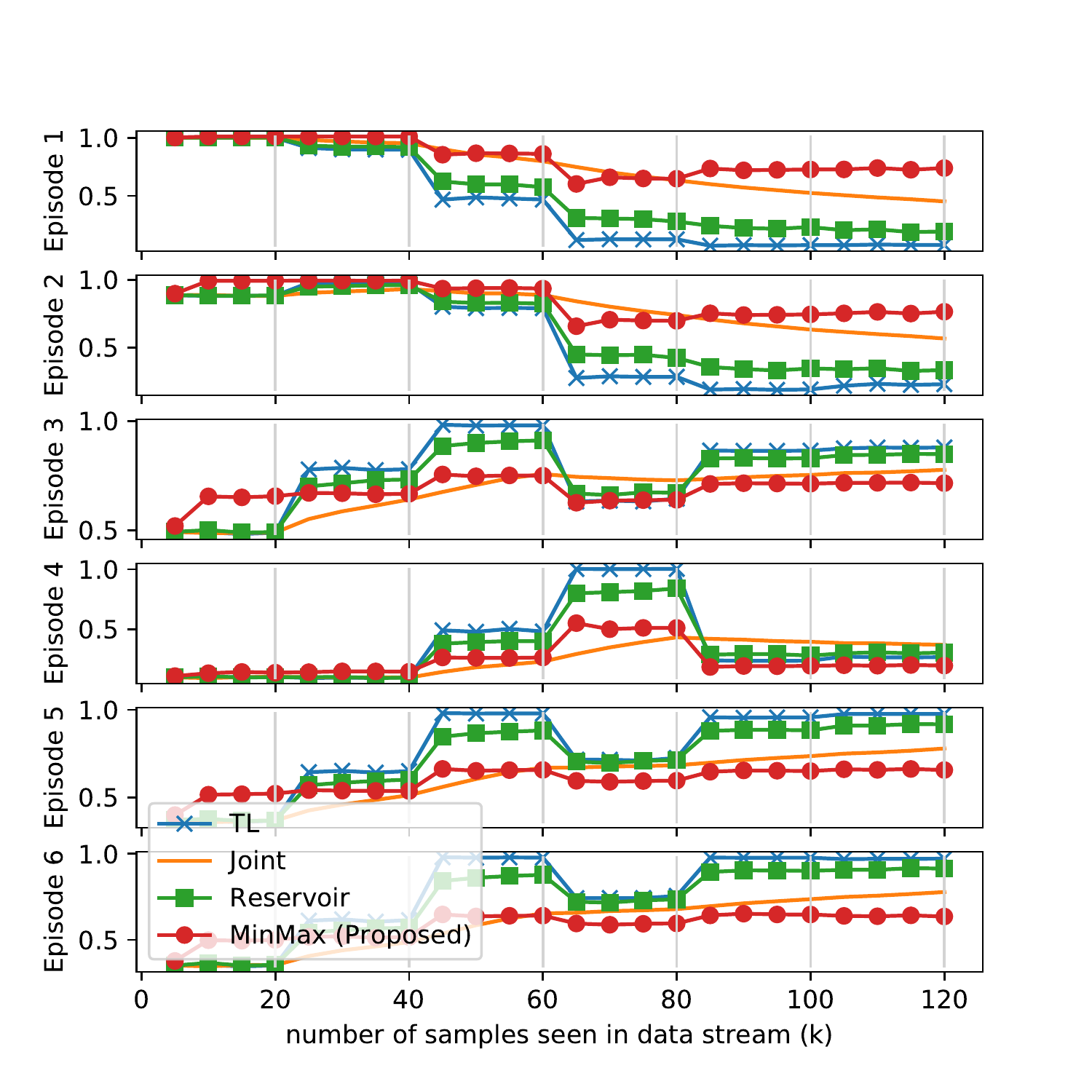}
\end{minipage}
\caption{Average sum-rate approximation ratios comparison on real measured channels. (left) performance on mixture of all six episodes. (right) per-task  performance. 
}
\label{ratio_pertask_mimo}
\end{figure}

\begin{figure}
\centering
\begin{minipage}[c]{0.4\linewidth}
    \centering
    \includegraphics[width=\textwidth]{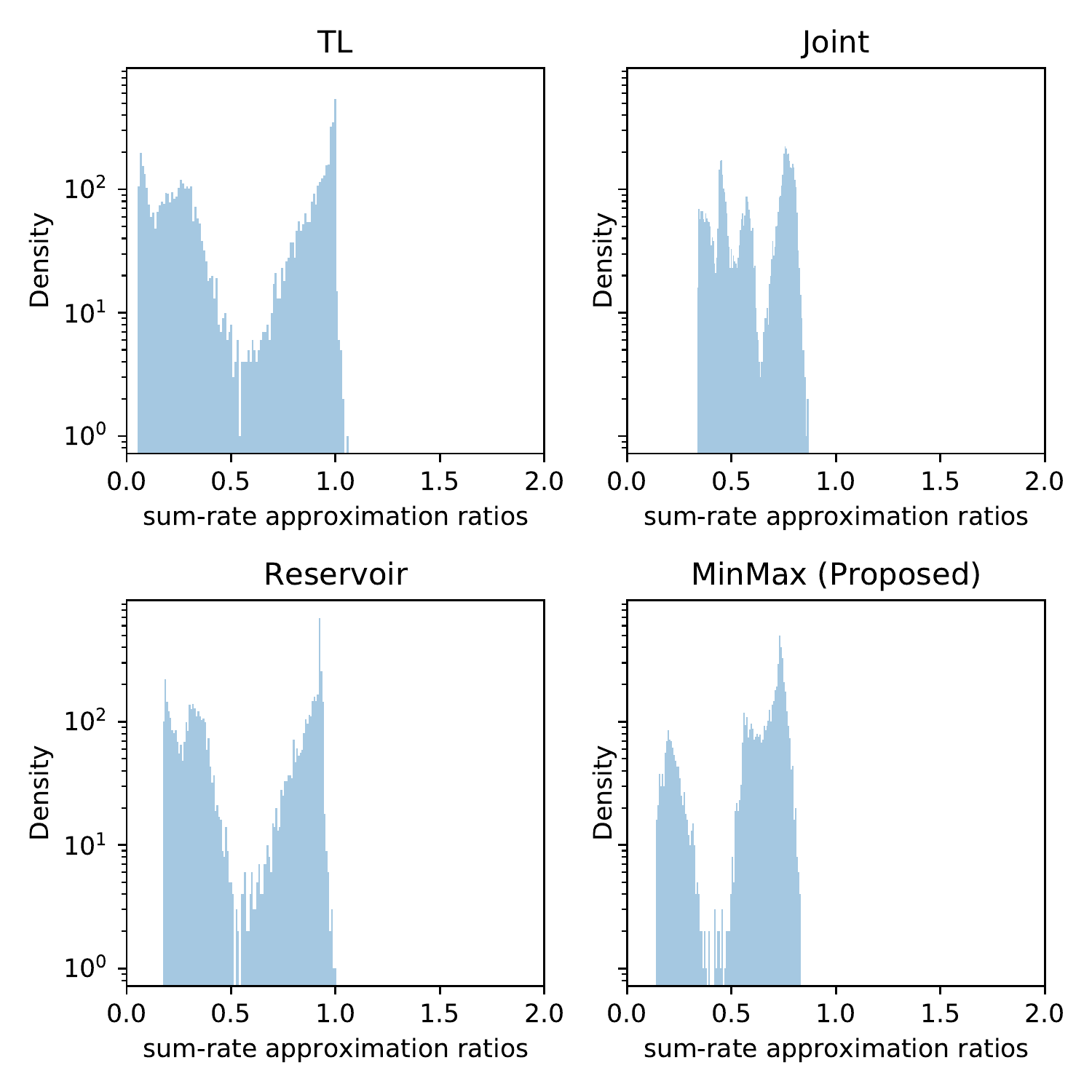}
    \footnotesize{(a) Probability Density Functions (PDF)}
\end{minipage}
\noindent
\begin{minipage}[c]{0.4\linewidth}
    \centering
    \includegraphics[width=\textwidth]{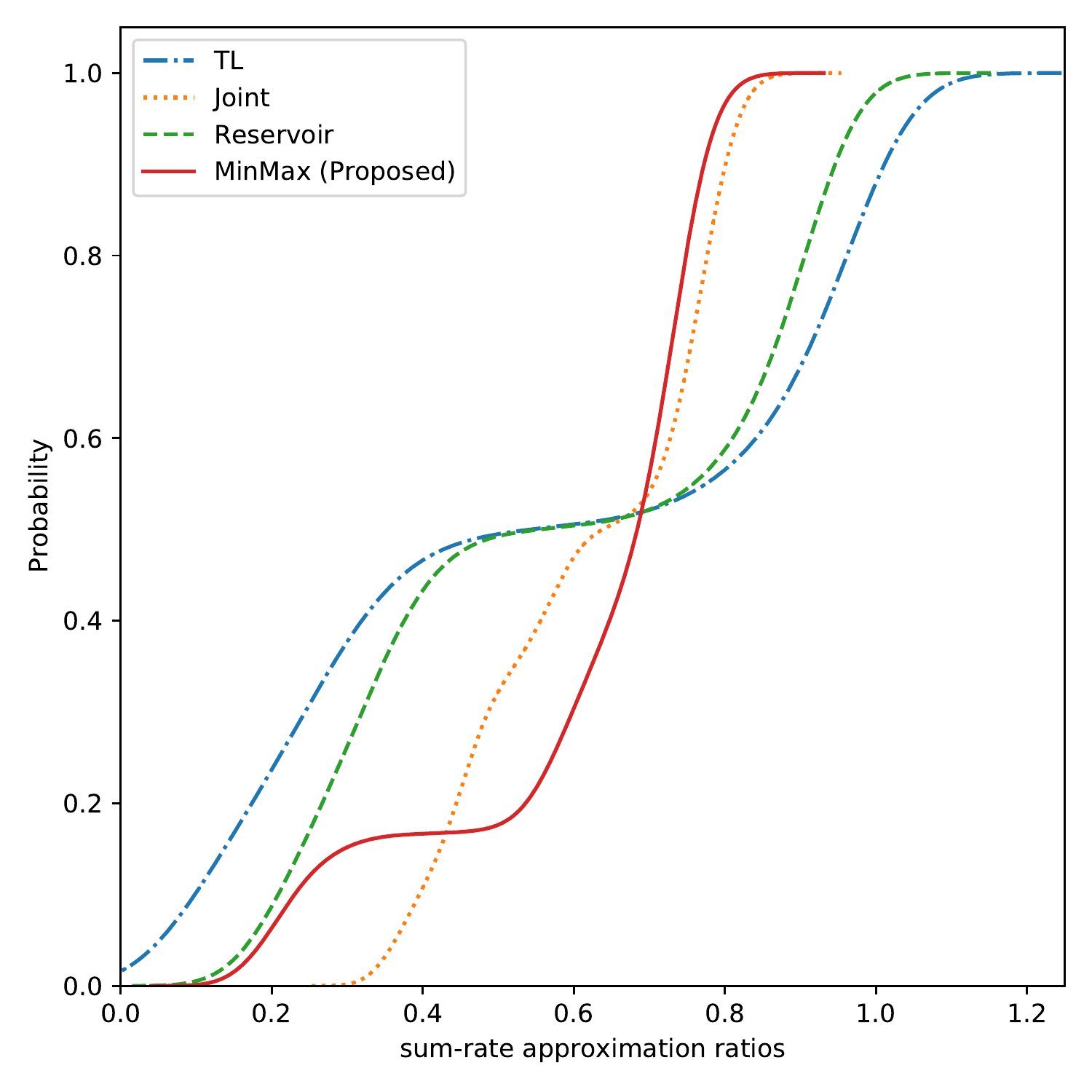}
    \footnotesize{(b) Cumulative Distribution Function (CDF)}
\end{minipage}
\caption{Distribution of the per-sample sum-rate  ratio evaluated at the last time stamp on the mixture test set of all six episodes.  }
\label{rate_dist_mimo}
\end{figure}

To validate our approach under more realistic scenarios, we further consider the outdoor `O1' ray-tracing scenario generated from the DeepMIMO dataset \cite{Alkhateeb2019}.
 The used dataset consists of two streets and one intersection, with the top-view showed in Fig. \ref{location}. The horizontal street is 600m long and 40m wide, and the vertical street is 440m long and 40m wide. We consider six different episodes that users gathering locations (i.e., red rectangles in Fig. \ref{location}) gradually shift following the order indicated by the numbers. For each episode, we generate $20,000$ channel realizations for training and $1,000$ for testing. For each channel realization, we generate the channel based on the same set of BSs $1$ to $10$ (i.e., blue circles in Fig. \ref{location}), and randomly pick $K = 10$ user locations from the indicated user area $i$ (i.e., red rectangles in Fig. \ref{location}). The BS is equipped with single antenna and has the maximum transmit power $p_k$=$30$dBm. The noise power is set to $-80$ dBm. The simulation results are shown in Fig.   \ref{rate_pertask_mimo} - \ref{rate_dist_mimo}.  From Fig.  \ref{rate_pertask_mimo}-\ref{ratio_pertask_mimo} we can observed that, after experiencing all six episodes, our proposed fairness based method is able to perform much better than transfer learning and reservoir sampling-based CL approach on the mixture test set.  This is because  both transfer learning and reservoir sampling-based CL approach suffer from forgetting issues for old episodes (i.e. episode 1 and 2), but the proposed fairness based method can perform well on episode 1 and 2 due to its fairness design -- focusing on under-performing episodes (i.e. episode 1 and 2) while relaxing on outperforming episodes (i.e. episode 5 and 6).  One interesting observation is that the proposed method is able to outperform the joint training with accumulated data in terms of the average sum-rate (cf. LHS of Fig.  \ref{rate_pertask_mimo}). This is because the joint training will treat all samples equally, thus if we have two episodes in opposite directions, the model will generate something in the middle. Instead, our proposed fairness based method will be more focus on one episode that generates the highest cost thus achieve higher average performance due to its fairness design.

\subsection{Beamforming Experiments} \label{sec:beamform}
Next, we further validate our CL based approach in the beamforming problems -- weighted sum-rate (WSR) maximization in multi-input and single-output (MISO) interference channels. Different from the previous sections where the deep learning model is based on the end-to-end learning \cite{sun2018learning}, we consider a deep unfolding based model proposed in \cite{zhu2020optimization} and \cite{zhu2020learning}. Specifically, by unfolding the parallel gradient projection (PGP) method, the parallel GP based recurrent neural network (RNN-PGP) method \cite{zhu2020learning} builds a recurrent neural network (RNN) architecture to learn the beamforming solutions efficiently. By leveraging the low-dimensional structures of the optimal beamforming solution, the constructed neural network can achieve high WSRs with significantly reduced runtime, while exhibiting favorable generalization capability with respect to the antenna number, BS number and the inter-BS distance. 

\begin{figure}
\centering
\begin{minipage}[c]{1\linewidth}
    \centering
    \includegraphics[width=0.8\textwidth]{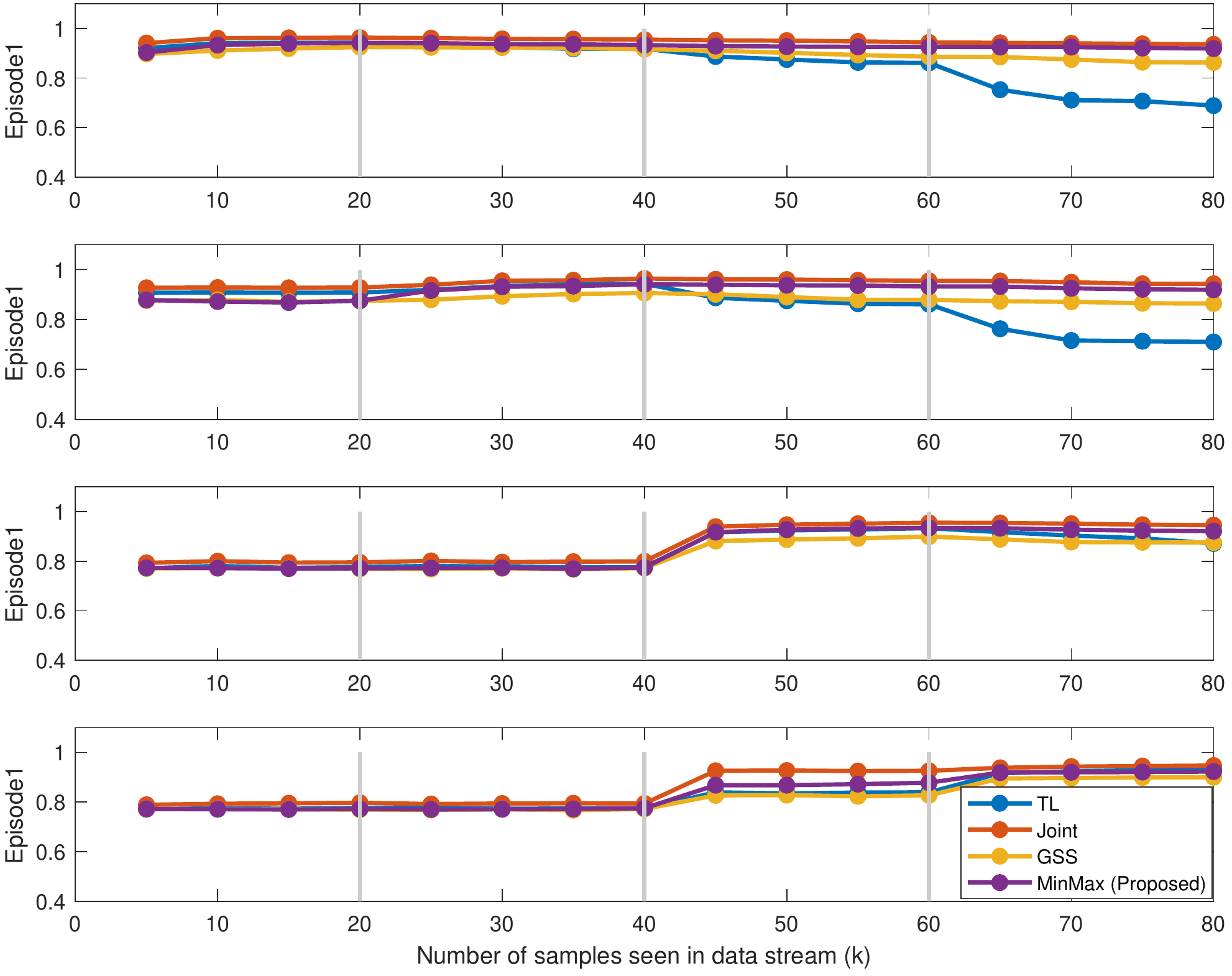}
    \caption{Achievable accuracy comparison for beamforming problem over the test set under randomly generated  channels, as the model learned progressively over time. }
    \label{beamform}
\end{minipage}
\end{figure}

In our simulation, we consider the same four types of channels described in Section \ref{subsection: Randomly Generated Channel}, but in the form of  cellular network described in \cite[Section 2]{zhu2020learning} with the number of base stations (BS) $K = 19$, number of neighbor considered $c =18$ and transmit antennas $N_t=36$. For the two episodes with geometry channels, we set the cell radius $d$ as $0.5$ km and $1$ km, respectively.  The simulation results over four different approaches are compared and reported in Fig. \ref{beamform}, where the x-axis represents the number of data samples has been observed, and the y-axis denotes the  ratio  of  the  (weighted)  sum-rate achieved by the RNN-PGP \cite{zhu2020learning} and that achieved by the WMMSE algorithm \cite{shi2011iteratively}. For our proposed approach (MinMax), we use the minmax formulation \eqref{minmax-ep} where both the training loss $\ell(\cdot)$ and the system performance loss $g(\cdot)$ are chosen as the MSE loss \eqref{MSE-loss}. It can be observed that the proposed algorithm (MinMax) almost matches the joint training performances, with only limited memory usage, and it  outperforms the GSS   approach. On the other hand, the naive transfer learning can  suffer from significant performance loss on old episodes (Episode 1 and 2 in our case).

\section{Conclusion and Future Works} \label{conclusion}

In this work, we design a new ``learning to continuously optimize'' framework for optimizing wireless resources in dynamic environments, where parameters such as CSIs keep changing. By building the notion of continual learning (CL) into the modeling process, our framework is able to seamlessly  and  efficiently  adapt  to  the  episodically dynamic  environment,  without  knowing  the  episode boundary, and most importantly, maintain high performance over all the previously encountered scenarios. 
The proposed approach is validated through two popular wireless resource allocation problems (one for power control and one for beamforming), two popular DNN based models (one for end-to-end learning and one for deep unfolding), and uses both synthetic and real data sets. Our empirical results make us believe  that our approaches can be extended to many other related problems.

Our work represents a preliminary step towards understanding the capability of deep learning for wireless problems with dynamic environments. There are many interesting questions to be addressed in the future, and some of them are listed below:
\begin{itemize}
    \item Is it possible to design  continual learning strategy  with theoretical guarantees?
    \item Is it possible to design theoretical results for the proposed stage-wise bi-level optimization problem \eqref{bilevel-ep} or even the global bi-level optimization problem \eqref{eq:bilevel}?
    \item Is it possible to quantify the generalization performance of the proposed fairness framework?
    \item Is it possible to extend our frameworks to other wireless tasks such as signal detection, channel estimation and CSI compression?
\end{itemize}

\newpage
\bibliographystyle{IEEEtran}
\bibliography{CL,WC,ref}

\end{document}